\documentclass{raa}

\usepackage{natbib}
\usepackage{graphicx}
\usepackage{epstopdf}

\usepackage{float}
\usepackage[pdftex,breaklinks]{hyperref}

\usepackage{savesym}
\usepackage{amsmath}
\savesymbol{iint}
\usepackage{txfonts}
\restoresymbol{TXF}{iint}

\usepackage[nointegrals]{wasysym}
\usepackage{etoolbox}

\bibpunct{(}{)}{;}{a}{}{,}

\begin{document}

\title{On the observed clustering of major bodies in solar and extrasolar subsystems}

   \volnopage{Vol.0 (20xx) No.0, 000--000}      
   \setcounter{page}{1}          

\author{
Dimitris M. Christodoulou\inst{1}  
\and 
Demosthenes Kazanas\inst{2}
}

\institute{
Lowell Center for Space Science and Technology, University of Massachusetts Lowell, \hfill\break Lowell, MA, 01854, USA; {\it dimitris\_christodoulou@uml.edu}\\
\and
NASA Goddard Space Flight Center, Laboratory for High-Energy Astrophysics, Code 663, Greenbelt, MD 20771, USA; {\it demos.kazanas@nasa.gov} \\
\vs\no
   {\small Received~~20xx month day; accepted~~20xx~~month day}}

\abstract{
Major (exo)planetary and satellite bodies seem to concentrate at intermediate areas of the radial distributions of all the objects present in each (sub)system. We prove rigorously that the secular evolution of (exo)planets and satellites necessarily results in the observed intermediate accumulation of the massive objects in all such subsystems. We quantify a ``middle'' as the mean of mean motions (orbital angular velocities) of three or more massive objects involved. Orbital evolution is expected to be halted or severely diminished when the survivors settle near mean-motion resonances and substantial angular-momentum transfer between bodies ceases to occur (gravitational Landau damping). The dynamics is opposite in direction to what has been theorized for viscous and magnetized accretion disks in which gas spreads out and away from either side of any conceivable intermediate area. The results are bound to change the way we think about planet and moon formation and evolution.
}

\authorrunning{Christodoulou and Kazanas}
\titlerunning{Major-body clustering in (extra)solar subsystems}

\maketitle

\section{Introduction}\label{intro}

An inspection of planetary and satellite orbital data in the solar system\footnote{{\tt https://ssd.jpl.nasa.gov}, {\tt https://solarsystem.nasa.gov}\label{ft1}} reveals that major objects seem to cluster at intermediate areas of the radial distributions of orbiting bodies, and only smaller objects are found in the inner and the outer regions of these subsystems. The same arrangement of massive objects is also seen in multiplanet extrasolar systems. Keeping in mind that there may be more undetected planets farther out in these systems, some examples presently are: HD 10180 \citep{lov11}, Kepler-80 and 90 \citep{sha18}, TRAPPIST-1 \citep{del18,gri18}, HR 8832 \citep{vog15,joh16,bon20}, K2-138 \citep{chrsen18,lop19}, Kepler-11 \citep{lis11}, and even the four-planet systems of Kepler-223 \citep{mil16} and GJ 876 \citep{riv10,mil18}. Despite being a clue pertaining to the processes of massive planet and satellite formation and evolution, this conspicuous property has not been discussed in the past, and there have been no ideas about how we could possibly exploit it to learn from it.

Our approach to the problem has been single-minded from the outset. It was apparent to us that such large bodies have moved toward one another during early evolution, perhaps as soon as a few large solid cores emerged in these subsystems and the accretion disks dissipated away. In such a case, there must exist a generic physical mechanism that drives this type of convergence but eventually further migration is hindered when the mechanism ceases to operate. In this work, we formulate such a secular mechanism that relies on first principles and requires no additional conditions in order to operate. Some related calculations have been carried out by other researchers in the past \citep{ost69,pag74,lyn74,bal98,pap11}. Any small differences that we may point out concern the details of evolution and the physical interpretation of the results. 

In \S~\ref{mri}, we describe the dynamical evolution of two interacting Keplerian fluid elements through nonequilibrium states that leads to a runaway dynamical instability. This analysis is applicable to magnetized accretion disks, but not to planets and satellites in which the integral of circulation is not conserved (not even approximately; these systems are topologically not simply-connected) precluding dynamical instability. In \S~\ref{analysis}, we describe the secular evolution of large individual gravitating bodies in Keplerian orbits around a central mass and under the influence of dissipation which leads to clustering of the bodies. In \S~\ref{dis}, we discuss our results in the context of planet and satellite evolution. 

Many technical details are left to three self-contained appendices. In Appendix~\ref{n3}, we describe few-body systems evolving by exchanging angular momentum and lowering their mechanical energies. In Appendix~\ref{app1}, we formulate a self-consistent calculation of the charactiristic dissipation time $\tau_{\rm dis}$ and the corresponding velocity fluctuations $v_{\rm dis}$ in such systems. In Appendix~\ref{alast}, we analyze ``gravitational Landau damping'' of the tidal field in few-body systems, a unique new mechanism that is responsible for settling of the bodies near mean-motion resonances over times comparable to $\tau_{\rm dis}$, where they no longer exchange substantial amounts of angular momentum and so they send the mean tidal field around them to oblivion.

\section{Dynamical Evolution of Keplerian Interacting Fluid Elements}\label{mri}

\cite{bal98} introduced a mechanical analog of the magnetorotational instability (MRI) in gaseous accretion disks, two mass elements $m_1$ and $m_2$ in circular Keplerian orbits around a central mass $M\gg m_1, m_2$ with radii $r_{1}$ and $r_{2} > r_1$, respectively. The mass elements are connected by a weak spring with constant $k$ (representing a magnetic-field line) whose role is to allow for angular momentum transfer between the elements. When perturbed under the constraint of constant total angular momentum,\footnote{Constant circulation would be more precise, although the two integrals of motion are equivalent in axisymmetric fluid systems.} this model behaves just like gaseous accretion disks under the influence of viscosity \citep{lyn74}, except that the instability is dynamical: the masses spread out and their displacements reduce the total free energy of the system \citep{chr95}, leading to a runaway \citep{bal91,bal98,chr96,chr03}.

We use the phase-transition formalism of \cite{chr95} to describe the evolution of this system {\it out of equilibrium}: a change that lowers the free energy ($\Delta E<0$) while preserving the total angular momentum ($\Delta L = 0$) is viable and the system will transition to the new nonequilibrium state of lower energy; whereas if $\Delta E > 0$, the system will just oscillate about the initial equilibrium state characterized by total energy $E=E_{1}+E_{2}$ and total angular momentum $L=L_{1}+L_{2}$. We assume that the initial equilibrium orbits are perturbed by small displacements $\Delta r_1 \ll r_{1}$ and $\Delta r_2 \ll r_{2}$. Then the conservation of total angular momentum relates the displacements to first order by the equation
\begin{equation}
m_1 v_{1}\Delta r_1 + m_2 v_{2}\Delta r_2 = 0\, ,
\label{xy}
\end{equation}
where $v_{1}$ and $v_{2}$ are the equilibium azimuthal velocities, and the change in free energy to first order is found to be
\begin{equation}
\Delta E = L_{1}(n_{1} - n_{2})\frac{\Delta r_1}{2 r_{1}}\, ,
\label{de}
\end{equation}
where $n_1$ and $n_2 < n_1$ are the mean motions (orbital angular velocities) of the masses in their equilibrium state. The change in potential energy of the spring, $k(\Delta r_2 - \Delta r_1)^2/2$, is of second order and is omitted from equation~(\ref{de}). It is now apparent that for $\Delta r_1 < 0$, then $\Delta E < 0$ and $\Delta r_2 > 0$. The masses spread out and the resulting nonequilibrium configuration is unstable to more spreading that reduces further the free energy of the system. 

The above dynamical instability (an analog of the MRI) does not operate in planetary and satellite systems. It is strictly applicable to perfect fluids in which circulation and angular momentum are both conserved \citep[as in][]{chr95}. Conservation of circulation is implicit in the above model; it can be readily seen in equation~(\ref{xy}) assuming that the mass elements are axisymmetric rings with equal masses, in which case the equation takes the form
\begin{equation}
v_{1}\Delta r_1 + v_{2}\Delta r_2 = 0\, ,
\label{xy2}
\end{equation}
to first order in the displacements.

In viscous unmagnetized disks, dissipative stresses destroy circulation slowly and the instability is then secular \citep[as in][]{lyn74}. In stellar and particle systems, there is no conservation law of circulation and equation~(\ref{xy2}) is invalid, even in approximate form, because all the elements of the stress tensor introduce gradients of comparable magnitude to the Jeans equations of motion \citep{bin87,chr95,bat00}. Therefore, the evolution of multiple planetary and satellite bodies requires a different mathematical approach, though still constrained by the applicable conservation laws of energy and angular momentum.

\section{Secular Evolution of Interacting Planets and Satellites}\label{analysis}

\cite{ost69} studied the secular evolution of a dynamically stable, uniformly-rotating pulsar subject to angular momentum and energy losses due to emission of multipolar radiation. Evolution takes place slowly over timescales much longer than the dynamical time (the rotation period) of the object. In this model, the pulsar is thought of as transitioning between quasistatic equilibrium states \citep[the Dedekind ellipsoids;][]{cha69} in which it maintains its uniform rotation albeit with a slowly changing angular velocity $\Omega$. Here, ``slowly'' is quantified by the condition that
\begin{equation}
\left\lvert\frac{d\Omega}{dt}\right\rvert \ll \Omega^2\, .
\label{cond}
\end{equation}
Under a series of assumptions, the strongest of which is inequality~(\ref{cond}), \cite{ost69} proved that the losses in angular momentum $L$ and kinetic energy $E$ are related by the equation
\begin{equation}
\frac{dE}{dt} = \Omega \frac{dL}{dt}\, ,
\label{dedt}
\end{equation}
where the time derivatives are both implicitly negative. The use of $E$ for rotational kinetic energy (their equation (7)) has caused some indiscretions in the literature. For example, \cite{pag74} call $E$ the ``energy-at-infinity'' (which is kinetic after all) and equation~(\ref{dedt}) ``universal'' despite having derived it under their assumption iv(a) which is essentially equivalent to inequality~(\ref{cond}); whereas \cite{pap11} treated $E$ as the mechanical energy of an orbiting planet within the same quasistatic approximation.

Below we also use equation~(\ref{dedt}) to follow the secular evolution of planets and satellites losing kinetic energy slowly due to the action of dissipative processes induced by the central object. First we revisit the approach of \cite{pap11} whose calculation is correct but his conclusion is wrong. Then we formulate the same problem as a variation of the free energy of the system undergoing quasistatic {\it out-of-equilibrium} evolution away from its initial equilibrium state.

\subsection{Papaloizou Approach}\label{pap}

We consider two gravitating bodies with masses $m_1$ and $m_2$ orbiting around a central mass $M\gg m_1, m_2$ in nearly circular Keplerian orbits with radii $r_1$ and $r_2 > r_1$, respectively. We assume that tides due to $M$ during orbit circularization are dissipated in the interiors of the bodies, causing small amounts of kinetic energy to be converted to heat $H$. The slow rate of dissipation is given by
\begin{equation}
{\cal L} = dH/dt > 0\, .
\label{dq}
\end{equation}
Here, ``slow'' is defined by inequality~(\ref{cond}) and by the condition that
\begin{equation}
H \ll T\, ,
\label{condt}
\end{equation}
where $T$ is the total kinetic energy. Then the evolution of the system is described by a sequence of {\it quasistatic equilibrium states} that are accessible to the bodies because equation~(\ref{dq}) along with energy conservation guarantee that the total mechanical energy of the bodies will decrease in time ($dE/dt < 0$).

The mechanical energy and angular momentum contents of each body are related by
\begin{equation}
E_i = -\frac{1}{2} n_i L_i\, ,
\label{el1}
\end{equation}
where $i=1, 2$ and $n_i$ is the mean motion of body $m_i$. Since $r_2 > r_1$, then $n_2 < n_1$ for the Keplerian orbits. Equation~(\ref{dedt}) is also valid here; under the quasistatic assumption~(\ref{cond}), it takes the form
\begin{equation}
\frac{dE_i}{dt} = -\frac{1}{2} n_i \frac{dL_i}{dt}\, .
\label{el2}
\end{equation}
The factor of $-1/2$ appears because $E_i$ represents the mechanical energy of each body which is implicitly negative. The negative sign cannot be absorbed in equations~(\ref{el1}) and~(\ref{el2}) because, unlike $dL/dt<0$ in equation~(\ref{dedt}) above, here the terms $dL_1/dt$ and $dL_2/dt$ have opposite signs.

Conservation of total angular momentum $L=L_1+L_2$ is expressed by the equation
\begin{equation}
\frac{d}{dt}\left(L_1 + L_2\right) = 0\, ,
\label{conl}
\end{equation}
and total energy conservation for the system gives
\begin{equation}
\frac{d}{dt}\left(E_1 + E_2\right) = -\frac{dH}{dt} = -{\cal L} < 0\, .
\label{cone}
\end{equation}
Using equations~(\ref{el2}), we rewrite equation~(\ref{conl}) in the form
\begin{equation}
\frac{1}{n_1}\frac{dE_1}{dt} + \frac{1}{n_2}\frac{dE_2}{dt} = 0\, .
\label{conl2}
\end{equation}
Thus, after considerable deliberations of the details, we have arrived at the equations adopted by \cite{pap11}. 

It is obvious from equations~(\ref{cone}) and~(\ref{conl2}) that, as the system evolves quasistatically, the mechanical energy of one body will increase and that of the other body will decrease, but the overall change in $E_1+E_2$ will be a decrease by an amount of $dH$, allowing for the system to proceed to a neighboring quasistatic equilibrium state. But it is not prudent to solve these equations for the energy rates in order to deduce the details of the evolution. It is more sensible to look at the changes in angular momentum of the bodies: Combining equations~(\ref{el2})-(\ref{cone}), we find that
\begin{equation}
 -\frac{dL_2}{dt} = \frac{dL_1}{dt} = \frac{2{\cal L}}{n_1 - n_2} > 0\, ,
\label{dL12}
\end{equation}
where ${\cal L} > 0$ and $n_1 > n_2$. We see now that the inner body 1 will gain angular momentum and will move outward, while the outer body 2 will lose angular momentum and will move inward. Overall, the two bodies will converge toward a common orbit in which they will share the total angular momentum equally. But in larger systems with 3 or more bodies, this convergence does not materialize once two-body interactions set in and such a common orbit proves to not be as important; especially since another critical orbit emerges characterized by the mean $\overline{n}$ of the mean motions $n_i$ of the bodies (see Appendix~\ref{n3}). For 3 or more bodies, this orbit is secularly unstable due to two- and three-body encounters between near-neighbors, but a body may remain in it for a long time, provided that another body does not come close. The significance and the repercussions of these results will be discussed in \S~\ref{dis} below.

\subsection{Free-Energy Variation Approach}\label{free}

Here we formulate the problem studied in \S~\ref{pap} as a variation of the free energy of the system of two bodies with masses $m_i$ ($i=1, 2$) orbiting around a central mass $M\gg m_i$ and stepping out of equilibrium and into a new state while still obeying conditions~(\ref{cond}), (\ref{condt}), and~(\ref{el2}). The two bodies can proceed to such a (generally nonequilibrium) state only if this state is characterized by lower free energy ($\Delta E<0$) and the same total angular momentum ($\Delta L = 0$). The total mechanical energy $E_1+E_2$ plays the role of the free energy \citep{chr95}, thus we have
\begin{equation}
\Delta\left(E_1 + E_2\right) < 0\, ,
\label{e1}
\end{equation}
and
\begin{equation}
\Delta\left(L_1 + L_2\right) = 0\, .
\label{e2}
\end{equation}
Combining these two relations with equations~(\ref{el2}) in the form $\Delta E_i = -(1/2) n_i\Delta L_i$, we find that
\begin{equation}
\left(n_2-n_1\right)\Delta L_2 = \left(n_1-n_2\right)\Delta L_1 > 0\, .
\label{e3}
\end{equation}
For $n_1>n_2$ (implying that the initial orbital radii obey $r_1 < r_2$), we find that $\Delta L_1>0$ and $\Delta L_2<0$, respectively. Thus, in order for the system to begin its search for a new equilibrium state of lower free energy, the inner body $m_1$ will move out and the outer body $m_2$ will move in.

\section{Discussion}\label{dis}

We have used the conservation laws of energy and angular momentum to describe and contrast the dynamical evolution of two interacting mass elements in a gaseous disk and the secular evolution of planets and satellites. Both types of subsystems were assumed to exhibit Keplerian orbital profiles around a dominant central mass and to exchange angular momentum via weak torques. Evolution however takes different paths in these two circumstances and the reason is the (non)conservation of circulation. In perfect-fluid disks (\S~\ref{mri}), circulation is conserved and the mechanical analog of the MRI turns out to be a dynamical instability \citep[as was first described by][]{bal98}; whereas in (extra)solar multi-body subsystems (\S~\ref{analysis}), there is no analogous conservation law and dissipative evolution proceeds secularly via a sequence of quasistatic equilibrium configurations \citep{ost69} or via nonequilibrium states, both of which have progressively lower mechanical energy compared to the preceding state.  

Extending the analytical work of \cite{pap11} to more than 2 orbiting bodies, we demonstrate in Appendix~\ref{n3} that tidal dissipation induced by the central mass leads to clustering of many-body systems generally toward the mean $\overline{n}$ of their mean motions $n_i$ ($i=1, 2, \cdots, N$, where $N\geq 4$). On the other hand, $N=2$ or $N=3$ major bodies may try to converge toward a common orbit\footnote{The common orbit with $\overline{L}$ does not stand out in multiple-body systems because a body that may reach it first will soon move out as transfer of angular momentum continues on. Only $N=2$ bodies can approach this orbit synchronously.} characterized by the mean $\overline{L}$ of their angular momenta, except for the third body if it happens to be near the critical orbit with mean motion $\overline{n}$. Although secularly unstable, this critical orbit may host a massive body for a long time, at least comparable to the dissipation time $\tau_{\rm dis}$ that characterizes this part of the evolution of the system ($\tau_{\rm dis}$ is quantified in Appendix~\ref{app1}). A close encounter with another body can clear out the critical orbit, if the convergence of bodies continues unimpeded for a long enough time (Appendix~\ref{n3}). Convergence of bodies may seem surprising to the reader, but it did not come as a surprise to us. In fact, we anticipated such an outcome because we were impressed by observations of the radial distributions of bodies in solar subsystems and exoplanetary systems (\S~\ref{intro}); they all show an unmistaken clustering of several (4-7) massive bodies at intermediate orbital locations around the critical orbit with mean motion $\overline{n}$.

The next obvious question is, where and how does such clustering of bodies stop? After all, the observed massive planets and satellites seem to be currently on very long-lived, if not secularly stable, orbits and no pair appears to be close to merging into the same orbit. So the clustering process must be quelled somehow before the objects begin interacting strongly via close paired encounters. Although we do not have a complete answer yet, we believe that we are well on our way toward understanding the final stages of orbital evolution: The seminal paper of \cite{gol65} provided a substantial part of the answer long ago. \cite{gol65} showed that several ``special cases of commensurable mean motions [of satellites] are not disrupted by tidal forces.'' This means that when some of the more massive satellites of the gaseous giants reach near mean-motion resonances (MMRs), they do not exchange angular momentum efficiently any more, thus they maintain their orbital elements in long-lasting dynamical configurations (see also Appendix~\ref{alast} for gravitational Landau damping of the mean tidal field when massive bodies approach MMRs). 

The most massive body must play a crucial role in the above process because it is the one that evolves tidally slower than all the other bodies, so it must be the body that lays out the resonant structure (i.e., the potential minima; see Appendix~\ref{alast}) of the tidal field for the entire subsystem. When other massive bodies reach close to nearby MMRs, their further evolution is impeded because the most massive body does not affect them tidally any longer; and they also refrain from interacting with smaller bodies. In this setting, the tidal field is thus severely damped and the remaining lower-mass objects that are trying slowly to converge will also be hampered, either because they encounter MMRs or they are simply too far away from the resonating massive bodies. In the end, the entire system will appear to be stable (no more substantial imbalances from exchanges of angular momentum) with all of its members lying in or near MMRs and the mean tidal field erased since the major bodies no longer contribute to it. At present, this is what is actually observed in all (exo)planetary and satellite subsystems, although we have not been able to communicate the results of our meta-studies yet (Christodoulou \& Kazanas, in prep.). For this reason, we clarify here what we perceive differently in reference to the volumes of work carried out about MMRs up until now\footnote{Page ~{\tt https://en.wikipedia.org/wiki/Orbital\_resonance}\hfill\break contains a comprehensive, albeit empirical, summary of orbital MMRs along with a listing of hyperlinks to $\sim$100 professional citations.\label{ft2}} \citep{roy54,gol65,wis80,wis86,mur99,mor02,riv10,lis11,fab14,chrsen18}: We believe that multiple-body resonances are {\it not} a local phenomenon; principal MMRs are global in each system and their locations are determined by the most massive object that used to dominate the mean tidal field spread out across the entire (sub)system. In such a global layout, it is inappropriate to use the relative deviations of orbital elements from exact nearby MMRs and set arbitrary thresholds for objects to be or not to be in resonance. Though unfortunately, we recognize this to be the current state of affairs in studies of phase angles of local MMRs between near-neighbors; for example, no-one else currently believes that the Earth is in the 1:12 resonance of Jupiter because its orbital period is 4.2 days longer than the exact resonant value of 361.05 d; and its phase angle would have to circulate slowly relative to the phase of Jupiter, so the same pattern would only repeat once every 87 years (see also the section on ``coincidental near MMRs'' in the citation of footnote~\ref{ft2} for the same argument). This of course is the wrong way to think about global resonances in a tidal field that appears nowadays to be severely damped. We defer further discussion of this rather complicated issue to Appendix~\ref{alast}.

The main result of this work has ramifications beyond the particular systems that we study. The orbits of the planets and satellites that we have in mind all have Keplerian radial profiles. The Keplerian profile is just a special case of a power law, a profile with no critical or inflection points, which makes it simple but featureless. But now, the dynamics of multiple bodies evolving by applying torques and exchanging angular momentum has given us a critical point in this profile, the mean $\overline{n}$ of the mean motions $n_i$, or equivalently, the harmonic mean $\overline{P}$ of the orbital periods $P_i$ ($i = 1, 2, \cdots, N$, where we take $N\geq 4$). Given $\overline{P}$, the critical orbital radius can be determined from Kepler's third law. We note however that perhaps not many bodies may be found occupying the critical orbits in their subsystems because all bodies may have {\it a priori} circularized their orbits at or near MMRs (unless of course the critical orbit coincides with an MMR, in which case the chances of finding a body there improve considerably). 

Our planetary system and Jupiter's satellite subsystem each contain $N=4$ dominant adjacent orbiting bodies, the gaseous giant planets and the Galilean moons, respectively. For the gaseous giants, we find that 
$$\overline{P} = 29.36~{\rm yr} ~~({\rm whereas}~P_{\rm Sa}=29.46~{\rm yr}),$$
so Saturn has settled just wide of the critical orbit as we see it presently. For the Galilean moons, we find that at present
$$\overline{P} = 3.82~{\rm d} ~~({\rm whereas}~P_{\rm Eu}=3.55~{\rm d}),$$
so Europa was trapped into the renowned Laplace resonance and could not expand its orbit farther out. We did not include inner low-mass bodies in these estimates for an obvious reason; their fates were fully determined by weak tidal forces exerted on them by the distant massive bodies, so they can be viewed as passive receivers of tiny amounts of angular momentum having slowly worked their way outward and toward the common goal. The Earth, in particular, may have taken angular momentum from nearby Mars, preventing the outward movement of this tiny planet.

For the Earth, it is interesting to examine where our planet finally settled at the end of the orbital evolution of the gaseous giants: our planet is currently orbiting just wide of the 1:12 principal MMR of Jupiter (as already mentioned, its orbital period is only 4.2 d longer). It is not surprising that the planet could not get rid of a small amount of angular momentum and fall back into the MMR. During secular evolution, it was only gaining tiny amounts of angular momentum working its way outward toward the common goal. Such slightly wider orbits are observed in many exoplanets as well \citep{lis11,fab14}. Those inner ones with orbital periods shorter than $\overline{P}$ may be understood along the same line of reasoning \citep[but see also][]{lit12,bat13}.

In extrasolar systems, K2-138 \citep{chrsen18,lop19} presents a transparent example of a planet on a critical orbit. For the six planets known in this system, we find that
$\overline{P} = 5.385~{\rm d} ~~({\rm whereas}~P_{d}=5.405~{\rm d}),$ so planet $d$ is effectively occupying the critical orbit. All planets are near global MMRs as determined from the orbital period of the largest planet $e$. In order of increasing orbital periods, these are 2:7, 3:7, 2:3 1:1, 3:2, 5:1, for planets $b$-$g$, respectively. In planets $b$-$f$, all adjacent pairs have local period ratios $P_{i+1}/P_i\simeq$ 3/2 \citep{chrsen18}; and the outermost planet $g$ resides in a higher-order harmonic, i.e., $P_{g}/P_e\simeq (3/2)^4$. The resonant chain is global, though not fully packed. If it were fully packed, then no planet would occupy the critical orbit.

Another example with the critical orbit being occupied is the TRAPPIST-1 system with seven planets in a very compact configuration \citep[$r_{\rm max}=0.062$ AU;][]{del18,gri18}. We find that
$\overline{P} = P_{d} = 4.050~{\rm d},$
so planet $d$ is on the critical orbit. All planets are near global MMRs as determined from the orbital period of the largest planet $g$. In order of increasing orbital periods, these are 1:8, 1:5, 1:3, 1:2, 3:4, 1:1, 3:2, for planets $b$-$h$, respectively. More details on how such systems came to be are included in Appendix~\ref{alast}.


\appendix
\section{Angular Momentum Transfer Between Mupltiple Bodies}\label{n3}
\subsection{Three Bodies}

\begin{figure}
\begin{center}
    \leavevmode
      \includegraphics[trim=1.5cm 0.85cm 0cm 0.35cm, clip, angle=0,width=9 cm]{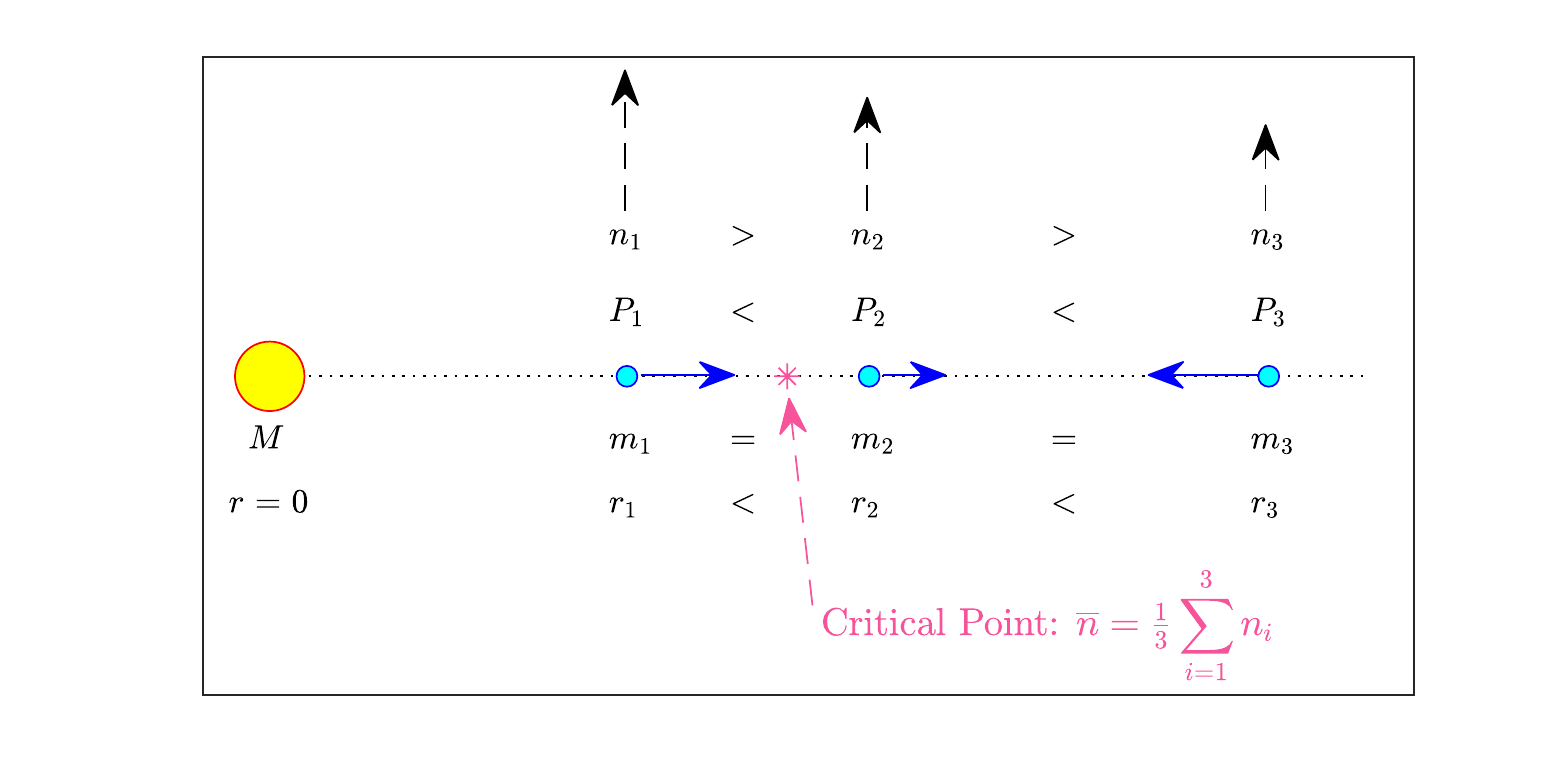}
\caption{Schematic diagram of three bodies in conjunction with equal masses $m_i$ ($i=1, 2, 3$) orbiting around (black arrows) a central mass $M\gg  m_i$ at radii $r_i$ with periods $P_i$. The asterisk denotes the location of the mean $\overline{n}$ of their mean motions $n_i$. The blue arrows indicate how the orbits will evolve initially via exchanges of angular momentum between the bodies.
\label{figa1}}
  \end{center}
\end{figure}

We consider three bodies with equal masses $m_i$ ($i=1, 2, 3$) orbiting around a cental mass $M\gg m_i$ in Keplerian orbits as illustrated schematically in Figure~\ref{figa1}. The mean motions $n_i = 2\pi/P_i$ obey the inequality $n_1 > n_2 > n_3$. We assume that the dissipation rate ${\cal L}>0$ is the same in all bodies and we use equation~(\ref{dL12}) to calculate the initial transfer of angular momentum $L_i\propto n_i^{-1/3}$ between pairs. We have
\begin{equation}
\frac{1}{2{\cal L}} \frac{dL_1}{dt} = \frac{1}{n_1-n_2} + \frac{1}{n_1-n_3}\, ,
\label{tr1}
\end{equation}
\begin{equation}
\frac{1}{2{\cal L}} \frac{dL_2}{dt} = \frac{-1}{n_1-n_2} + \frac{1}{n_2-n_3} 
\, ,
\label{tr2}
\end{equation}
and
\begin{equation}
\frac{1}{2{\cal L}} \frac{dL_3}{dt} = \frac{-1}{n_1-n_3} + \frac{-1}{n_2-n_3}
\, .
\label{tr3}
\end{equation}
The total orbital angular momentum of the system is indeed conserved; adding these three equations and simplifying, we obtain
\begin{equation}
\frac{d}{dt}\left(L_1+L_2+L_3\right) = 0
\, .
\label{tr4}
\end{equation}
Since $n_1 > n_2 > n_3$, then it becomes clear that $dL_1/dt>0$ and $dL_3/dt < 0$, so $m_1$ and $m_3$ will converge toward $m_2$. For body $m_2$, we rewrite equation~(\ref{tr2}) as
\begin{equation}
\frac{1}{2{\cal L}} \frac{dL_2}{dt} =\frac{n_1 - 2n_2 + n_3}{(n_1-n_2)(n_2-n_3)}
\, .
\label{tr5}
\end{equation}
We find that $dL_2/dt = 0$ if and only if $m_2$ is orbiting at the average value of the mean motions $n_1$ and $n_3$ (i.e., if $n_2 = (n_1+n_3)/2$), which, by a property of the arithmetic mean of a sequence of numbers, is also equal to
\begin{equation}
\overline{n} = \frac{1}{3}\left(n_1+n_2+n_3\right);
\label{meann}
\end{equation}
in such a case, $m_2$ facilitates the transfer of angular momentum between $m_1$ and $m_3$ without being subjected to a net gain or loss in $L_2$. In fact, $m_2$ acts as a forward-biased conduit that transfers angular momentum from $m_3$ to $m_1$.

\begin{figure}
\begin{center}
    \leavevmode
      \includegraphics[trim=0.1cm 0.1cm 0.1cm 0.1cm, clip, angle=0,width=9 cm]{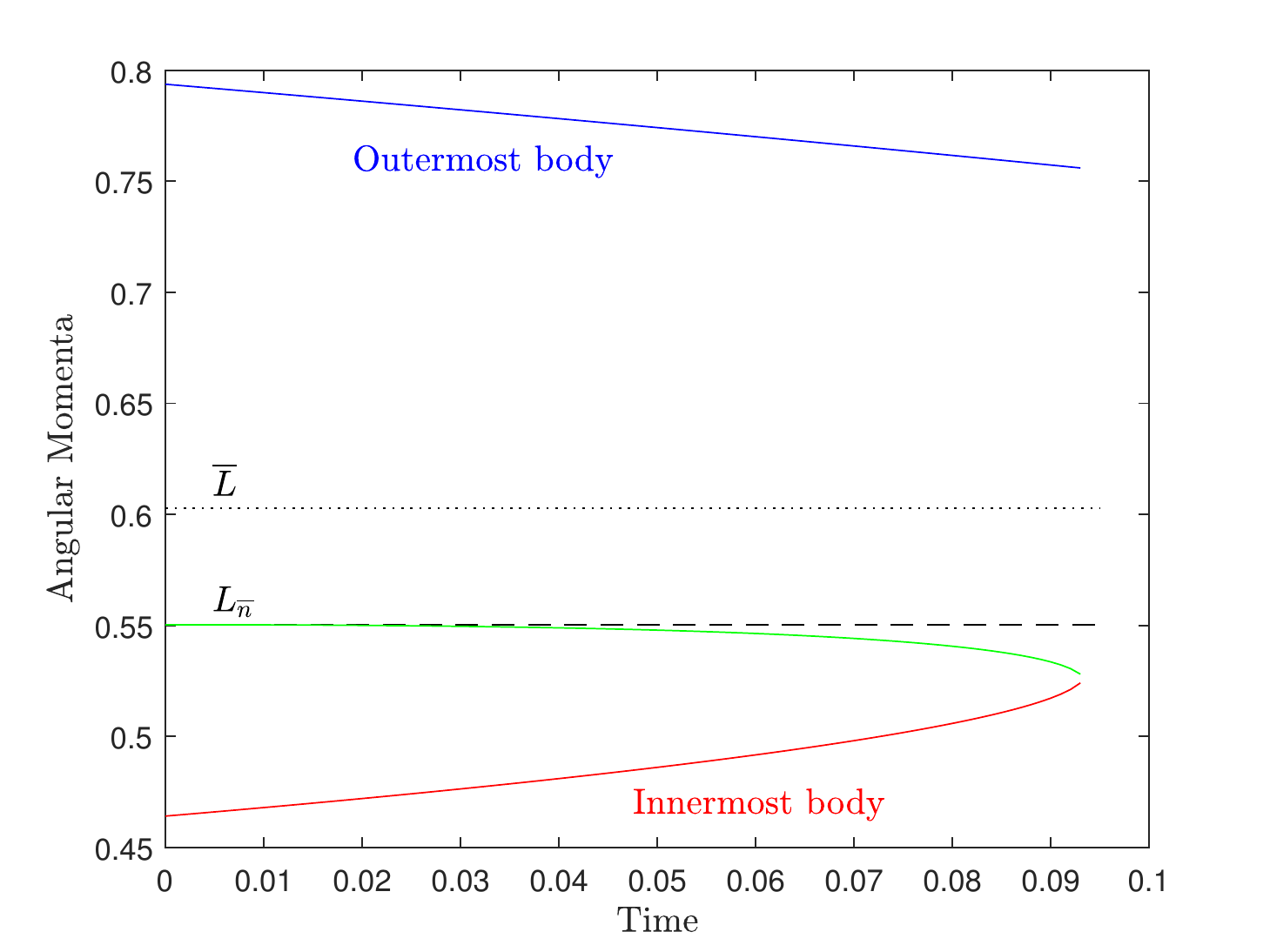}
\caption{Time evolution of the angular momenta of 3 equal-mass bodies with initial conditions $n_1=10$, $n_2=6$, $n_3=2$, $L_i=n_i^{-1/3}$ (Keplerian orbits), and $\overline{n} = (n_1+n_2+n_3)/3$. Time is measured in units of $QP$, where $Q$ is the effective tidal dissipation function and $P$ is the orbital period ($QP > \tau_{\rm dis}\sim Q^{1/2}P$; Appendix~\ref{app1}). This early evolution does not depend on the chosen timestep. Body 2 starts with $n_2=\overline{n}$ and $L_2=L_{\overline{n}} = (\overline{n})^{-1/3}$. The total angular momentum $L_{\rm tot}$ of the system is conserved and $\overline{L} = L_{\rm tot}/3$.
\label{figa2}}
  \end{center}
\end{figure}

Furthermore, if $n_2<\overline{n}$ (as shown in Figure~\ref{figa1}), then $dL_2/dt >0$ (equation~(\ref{tr5})) and the orbit of $m_2$ will expand; whereas the opposite will occur for $n_2>\overline{n}$. Thus, this critical orbit characterized by $\overline{n}$ is secularly unstable, but a body placed in it may survive for a long time and until it undergoes a close encounter with another approaching body. This is shown in Figure~\ref{figa2} that depicts the evolution of 3 equal-mass bodies in Keplerian orbits with initial conditions $n_1=10$, $n_2=\overline{n}=6$, and $n_3=2$.

\begin{figure}
\begin{center}
    \leavevmode
      \includegraphics[trim=0.1cm 0.1cm 0.1cm 0.1cm, clip, angle=0,width=9 cm]{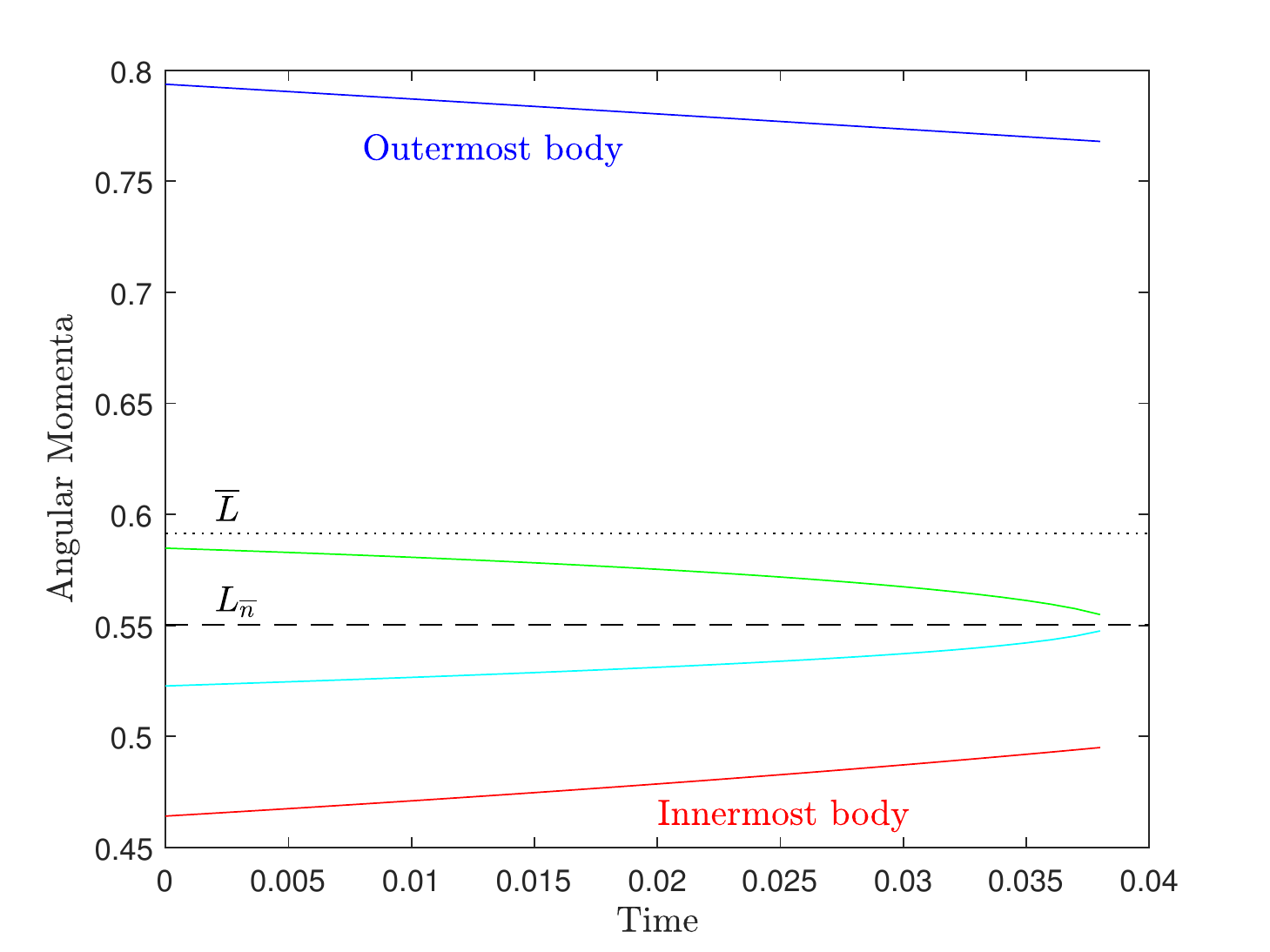}
\caption{As in Figure~\ref{figa2}, but for 4 equal-mass bodies with initial conditions $n_1=10$, $n_2=7$, $n_3=5$, $n_4=2$, and $\overline{n}=6$.
\label{figa3}}
  \end{center}
\end{figure}

\begin{figure}
\begin{center}
    \leavevmode
      \includegraphics[trim=0.1cm 0.1cm 0.1cm 0.1cm, clip, angle=0,width=9 cm]{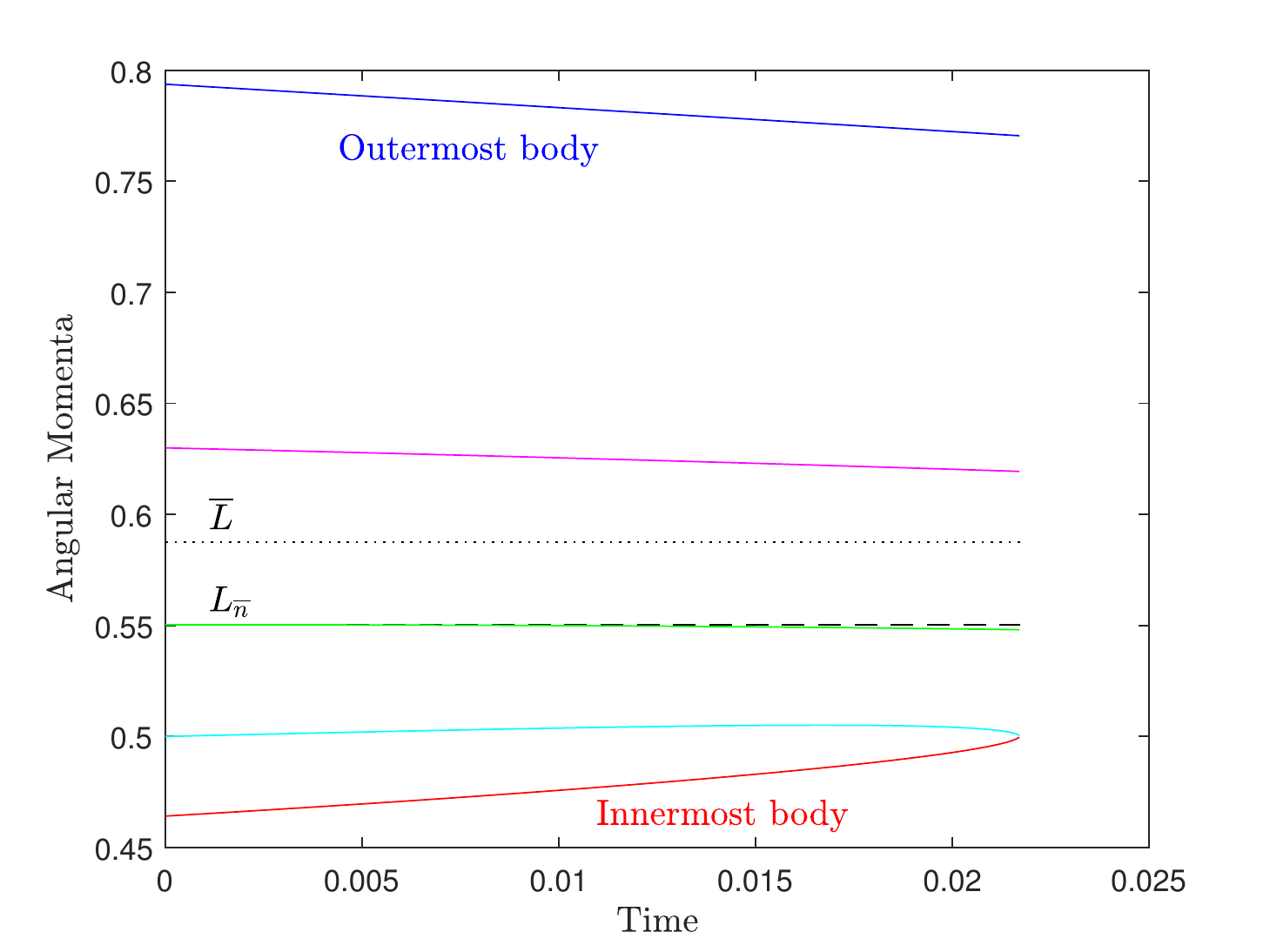}
\caption{As in Figure~\ref{figa2}, but for 5 equal-mass bodies with initial conditions $n_1=10$, $n_2=8$, $n^\prime=\overline{n}=6$, $n_3=4$, and $n_4=2$.
\label{figa4}}
  \end{center}
\end{figure}

\begin{figure}
\begin{center}
    \leavevmode
      \includegraphics[trim=0.1cm 0.1cm 0.1cm 0.1cm, clip, angle=0,width=9 cm]{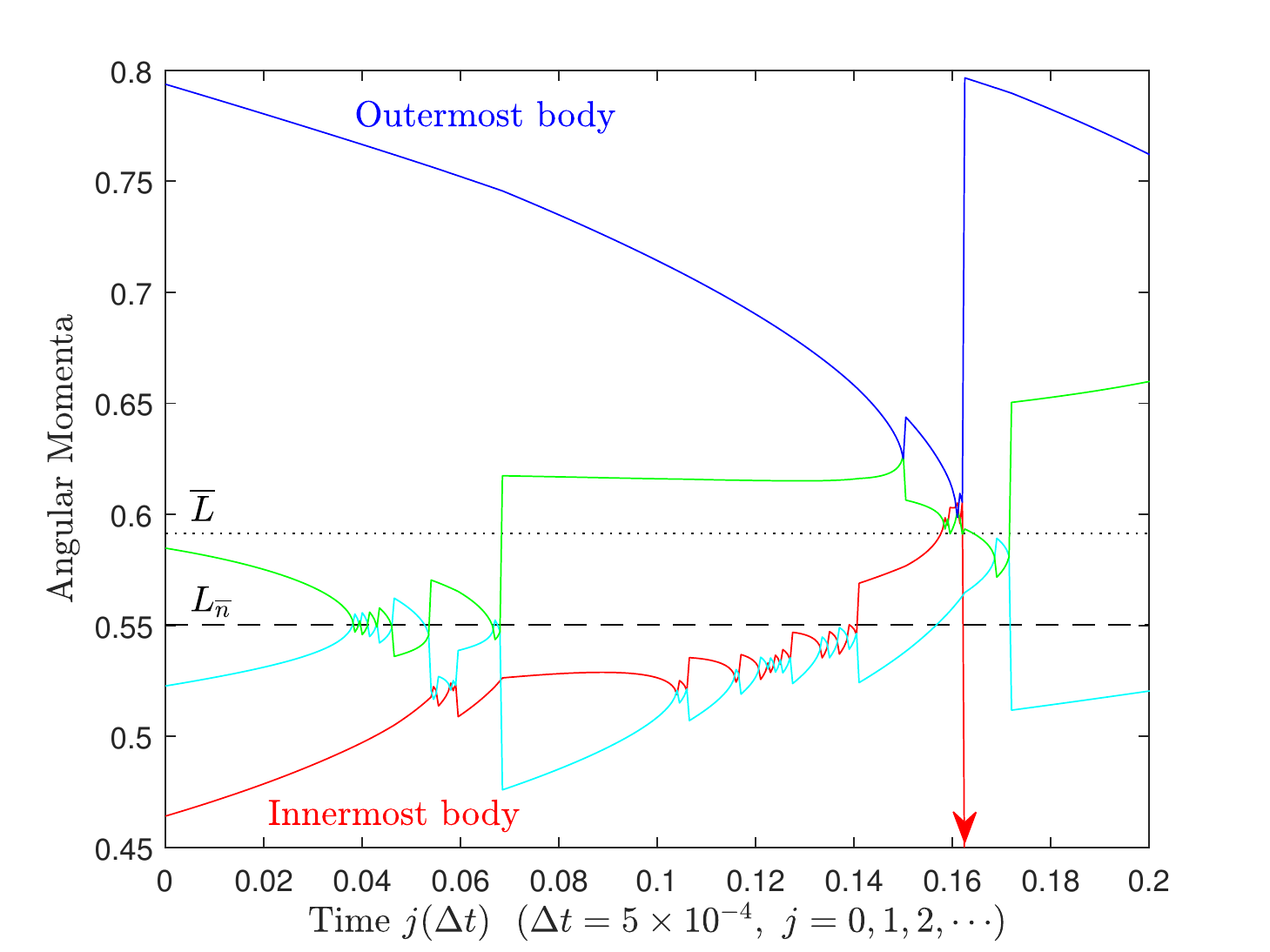}
\caption{Long-term ($5\times$) evolution of the angular momenta of the 4 equal-mass bodies shown in Figure~\ref{figa3}. Close encounters between pairs and triplets are dependent on the chosen timestep $\Delta t = 5\times 10^{-4}$.
\label{figa6}}
  \end{center}
\end{figure}

\begin{figure}
\begin{center}
    \leavevmode
      \includegraphics[trim=0.1cm 0.1cm 0.1cm 0.1cm, clip, angle=0,width=9 cm]{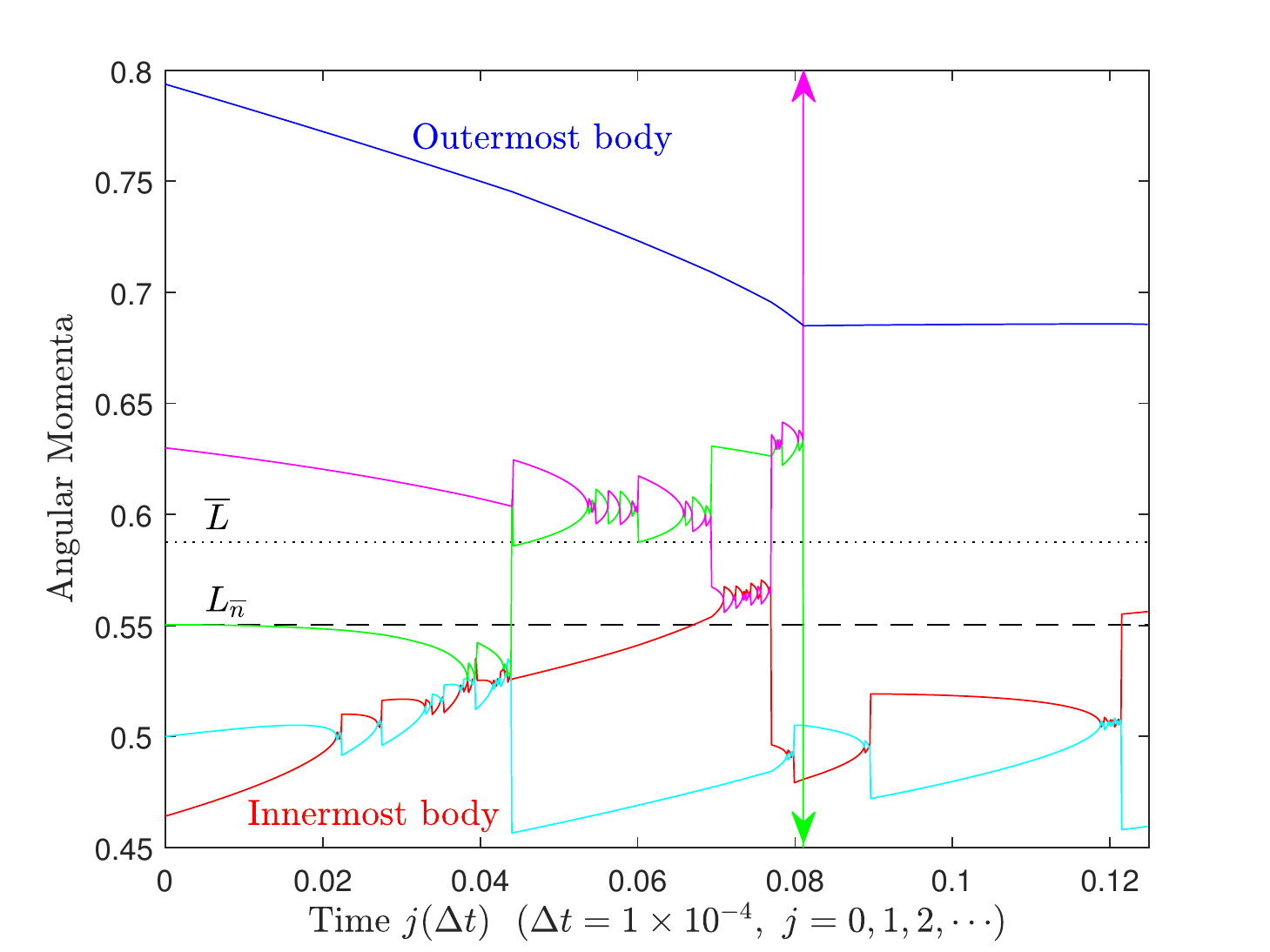}
\caption{Long-term ($5\times$) evolution of the angular momenta of the 5 equal-mass bodies shown in Figure~\ref{figa4}. Close encounters between pairs and triplets are dependent on the chosen timestep $\Delta t = 1\times 10^{-4}$.
\label{figalast}}
  \end{center}
\end{figure}

\subsection{Four Bodies}

As can be seen in Figures~\ref{figa3} and~\ref{figa4}, the same behavior and conclusions can be deduced for four (or more) equal-mass orbiting bodies for the times before paired interactions begin to occur. As one or two bodies may reach near the critical orbit with mean motion $\overline{n}$, the remaining bodies will continue exchanging smaller amounts of angular momentum. Left unimpeded, this process will lead to orbit coalescence (possibly after ejection of some closely interacting pairs; see Figures~\ref{figa6} and~\ref{figalast}), and this is why in (extra)solar subsystems there must be another mechanism to quell or severely depress angular momentum transfer before the orbits merge. As was discussed in \S~\ref{dis}, we believe that such a mechanism has been discovered by \cite{gol65} long ago.

Referring back to the results depicted in Figure~\ref{figa3}, it is important to investigate the early behavior of the two intermediate bodies (2 and 3) lying near the critical orbit with $\overline{n}$ in a four-body system, when paired interactions between nearest neighbors are not too strong yet. This is because the two most famous subsystems in our solar system, the gaseous giant planets and the Galilean satellites of Jupiter, both contain four major bodies each. In this case, we find for bodies 2 and 3 that
\begin{equation}
\frac{1}{2{\cal L}} \frac{dL_2}{dt} =\frac{n_1-2n_2+n_3}{(n_1-n_2)(n_2-n_3)} + \frac{1}{n_2-n_4}
\, ,
\label{2of4_1}
\end{equation}
and
\begin{equation}
\frac{1}{2{\cal L}} \frac{dL_3}{dt} = \frac{n_2-2n_3+n_4}{(n_2-n_3)(n_3-n_4)} - \frac{1}{n_1-n_3}
\, .
\label{3of4_1}
\end{equation}
We see that a sufficient condition for $dL_2/dt >0$ is that $n_2\leq (n_1+n_3)/2$; and a sufficient condition for $dL_3/dt<0$ is that $n_3\geq (n_2+n_4)/2$. In this case, bodies 2 and 3 will initially converge toward one another irrespective of the location of $\overline{n}$. In a variety of cases however, the four major bodies are expected to have formed at some relative distances from one another, and $\overline{n}$ may just as well have initially fallen between $n_2$ and $n_3$. Then the two intermediate bodies 2 and 3 will converge toward $\overline{n}$, as seen in Figure~\ref{figa3}.

\subsection{Five Bodies}

Here we investigate the critical orbit with $\overline{n}$ in the case of 5 interacting bodies. In a four-body configuration, we place initially a fifth body with $n^\prime$ and $L^\prime$ at the critical orbit of the other four bodies 1-4, so that 
\begin{equation}
n^\prime = \overline{n} = \frac{1}{4}\left(n_1+n_2+n_3+n_4\right)\, .
\label{mean5}
\end{equation}
The mean $\overline{n}$ remains unchanged for the 5 bodies. After some tedious algebra, the angular momentum change for the added body is found to be proportional to three cyclic factors, viz.
\begin{equation}
\frac{dL^\prime}{dt} \propto \left( n_1+n_2-2n^\prime\right) \left( n_2+n_3-2n^\prime \right) \left( n_1+n_3-2n^\prime \right) ,
\label{L5_1}
\end{equation}
where all positive definite factors have been dropped for the sake of convenience. We find that the initial condition $n^\prime = \overline{n}$ is not sufficient for the fifth body to be in an equilibrium orbit with $dL^\prime/dt=0$, but another condition must also be met. If $dL^\prime$ is set to zero, the above 3 factors determine the additional condition for $n^\prime$ to be the average of any of the specific 3 pairs of mean motions. Each of these averages is cyclically equivalent to yet another average between mean motions (1,2$\to$3,4; 2,3$\to$4,1; and 1,3$\to$4,2), for a total of 6 combinations between any two paired mean motions. The first two averages (1,2$\to$3,4) cannot occur, but the remaining four combinations are viable.

Any one of the four viable conditions, along with $n^\prime = \overline{n}$, is sufficient for the fifth body to be initially in equilibrium. Such an equilibrium state is unstable due to interaction of the fifth body with any other body that may come close in the long term. But this state can be long-lived if the nearest neighbors take a long time to approach the fifth body. An example of the entire process, complete with two- and three-body interactions at later times, is shown in Figures~\ref{figa4} and~\ref{figalast}, respectively, in which the initial setup of the 5 mean motions is symmetric about $\overline{n}=6$.

\subsection{Time Evolution of the Critical Orbit}

The only constant plotted in the figures above is the mean angular momentum $\overline{L}$. The corresponding mean motion $n_{\overline{L}}=(1/\overline{L})^{3}$ of such a ``common'' Keplerian orbit is also constant in time. This orbit is far less important for systems with 4 or more bodies and for 3 bodies one of which occupies initially the critical orbit (Figures~\ref{figa3}, \ref{figa4} and Figure~\ref{figa2}, respectively). On the other hand, the important critical orbit with initial values $\overline{n}$ and $L_{\overline{n}}$ does not remain constant in time; it relocates slowly toward the constant common orbit with $\overline{L}$. In fact, all orbits would do the same in the absence of close encounters during which large amounts of angular momentum are exchanged. Before any such encounters, a body placed initially on the critical orbit with ${\overline{n}}$ cannot regulate the transfer of angular momentum throughout the system to stay on this slowly-changing orbit; instead, it remains near its original orbit because its angular momentum content varies very slowly in time (Figure~\ref{figa4}).

As the critical orbit moves toward the constant common orbit, its angular momentum $L_{\overline{n}}$ always increases and its mean motion $\overline{n}$ always decreases in a Keplerian setting. That is, schematically,
\begin{equation}
L_{\overline{n}} ~~\overrightarrow{<}~~ \overline{L} = {\rm constant},
\label{mov1}
\end{equation}
and
\begin{equation}
\overline{n} ~~\overrightarrow{>}~~ n_{\overline{L}} = (1/\overline{L})^{3}= {\rm constant}.
\label{mov2}
\end{equation}
Proving one of these inequalities is not a trivial matter (the other one follows immediately for Keplerian rotation). With the help of {\it Mathematica}, we have shown that inequality~(\ref{mov2}) is an identity for $N=2$ and $N=3$ bodies, so an inductive proof may be possible although it does not appear to be mathematically tractable.

\section{Dissipation Timescale, Velocity Fluctuations, and Related Scales}\label{app1}

\subsection{Dissipation Timescale}

We estimate the dissipation timescale $\tau_{\rm dis}$ for interacting bodies such as massive planets and large  satellites. We begin with the Kolmogorov microscales, in which viscosity dominates and a small part of the kinetic energy is converted to heat. Although these microscales are used to describe diffusion in fluids, the equations are relevant to our problem as well because they imply a Reynolds number of $Re=1$ \citep{lan92}, a value that is appropriate for stellar systems and multiple bodies evolving quasistatically under the influence of weak tidal interactions. For $Re=1$, the square of the dissipation time is
\begin{equation}
\tau_{\rm dis}^2 \equiv \frac{\nu}{\epsilon}
\, ,
\label{a1}
\end{equation}
where $\nu$ is the kinematic viscosity coefficient and $\epsilon$ is the specific (per unit mass) energy dissipation rate. These quantities are related by $\epsilon = 2\nu\,\overline{e_{ij}e_{ij}}$, where $e_{ij} ~(i\neq j)$ is the symmetric strain-rate tensor that appears in the equations of motion \citep{bat00}. Thus, $\tau_{dis}$ in equation~(\ref{a1}) is the root mean square value of the reciprocal terms $1/e_{ij}$. This property allows us to estimate $\nu$ and $\epsilon$ from the macroscopic scales of interest\footnote{That the specific energy dissipation rate $\epsilon$ is determined over much larger (macroscopic) length scales $l$ than the turbulent dissipative microscales $\delta_{\rm dis}$ is well-known in studies of turbulent fluids \citep[see \S~\ref{IL} below and][]{geo13}.} (one cycle with orbital period $P$) without altering the microscopic gradients of the strains that literally do all the work. For the viscosity coefficient $\nu$ with dimensions of area over time, we write for one cycle that
\begin{equation}
\nu = \frac{\pi r^2}{P}
\, ,
\label{a2}
\end{equation}
where $r$ and $P$ are the orbital radius and orbital period, respectively; and for the specific energy dissipation rate $\epsilon$ with dimensions of power per unit mass, we write
\begin{equation}
\epsilon = \frac{1}{\mu}\left(-\frac{dE}{dt}\right) \equiv \frac{{\cal L}}{\mu}
\, ,
\label{a3}
\end{equation}
where ${\cal L}>0$ and $\mu$ is the distorted mass in which dissipation occurs in each cycle (i.e., the mass of the tidal bulges in a body). Combining equations~(\ref{a1})-(\ref{a3}), we find that
\begin{equation}
\tau_{\rm dis}^2 = \frac{\pi \mu r^2}{{\cal L}P}
\, .
\label{a4}
\end{equation}

We relate ${\cal L}$ to the effective specific tidal dissipation function $Q$ \citep{mun60,mac64,gol66} by estimating the kinetic energy loss of mass $\mu$ over one cycle $P$, viz.
\begin{equation}
{\cal L}P = \oint_P{\left(-\frac{dE}{dt}\right)dt} \equiv \frac{2\pi}{Q}T_0
\, ,
\label{a6}
\end{equation}
where $T_0=\mu\Omega^2 r^2/2$ is the orbital kinetic energy of mass $\mu$ and $Q \gg 1$ is a dimensionless function. Here we assume that the rotational kinetic energy $T_{\rm R}$ of $\mu$ is negligible compared to $T_0$. Equation~(\ref{a6}) implies that 
\begin{equation}
\epsilon = \frac{{\cal L}}{\mu} = \frac{\pi\Omega^2 r^2}{Q P}
\, ,
\label{a7}
\end{equation}
and substitution into equation~(\ref{a4}) gives 
\begin{equation}
\left(\Omega \tau_{\rm dis}\right)^2 = Q\, ,
\label{ot}
\end{equation}
or 
\begin{equation}
\tau_{\rm dis} = \frac{Q^{1/2}}{\Omega} = \frac{Q^{1/2}}{2\pi}P
\, ,
\label{a8}
\end{equation}
where we have used $\Omega\equiv 2\pi/P$. These equations may be useful for estimating dissipation times $\tau_{\rm dis}\gg P$ (where $Q\gg 1$), but they do not provide clear physical insight. For this reason, we recast equation~(\ref{a4}) in the form
\begin{equation}
\tau_{\rm dis}^2 = \frac{L}{2{\cal L}}
\, ,
\label{a10}
\end{equation}
where $L=\mu\Omega r^2$ is the total angular momentum of mass $\mu$, and for an average energy dissipation rate of ${\cal L} = \Delta E/\tau_{\rm dis}$, we find that
\begin{equation}
\tau_{\rm dis} = \frac{L}{2\,\Delta E} = \frac{\ell}{2\,\Delta\varepsilon}
\, ,
\label{a11}
\end{equation}
where $\ell = L/\mu$ and $\Delta\varepsilon = \Delta E/\mu$ are the corresponding specific quantities, respectively. We see now that $\tau_{\rm dis}$ is the time it takes to dissipate a part $\Delta E$ of the energy at constant bulge angular momentum $L$; or in microscales, the time to dissipate a part $\Delta\varepsilon$ of the specific energy at constant specific angular momentum $\ell$. 

Equation~(\ref{a11}) can also be recast in the familiar form
\begin{equation}
\Delta E = \frac{1}{2}I\overline{\omega}^2
\, ,
\label{a12a} 
\end{equation}
where $I=L/\Omega$ is the orbital moment of inertia of mass $\mu$ and 
\begin{equation}
\overline{\omega} \equiv \sqrt{\Omega\,(\tau_{\rm dis})^{-1}}
\, ,
\label{a12b}
\end{equation}
is the geometric mean of the two characteristic frequencies of the problem. Equation~(\ref{a12a}) justifies the presence of the factor of 1/2 in equations~(\ref{a10}) and~(\ref{a11}) above; whereas equation~(\ref{a12b}) shows how the dissipation couples to orbital dynamics and regulates the energy loss $\Delta E$ of the tidal bulges during quasistatic evolution. Clearly, the geometric mean $\overline{\omega}$ places more weight to $(\tau_{\rm dis})^{-1}$, the much shorter one of the two frequencies. This is seen also in the equivalent relation $\overline{\omega} = \Omega/Q^{1/4}$, where $Q\gg 1$ and $\overline{\omega} \ll \Omega$.

\subsection{Velocity Fluctuations}

We relate the tidal dissipation function $Q$ to the characteristic velocity $v_{\rm dis}$ of small-scale fluctutions which, for $Re=1$, is given by \citep{lan92} as
\begin{equation}
v_{\rm dis}^4 \equiv \nu \epsilon
\, .
\label{q1}
\end{equation}
Using equations~(\ref{a1}), (\ref{a7}), (\ref{a8}) and~(\ref{q1}), we find that
\begin{equation}
Q = \frac{1}{4}\left(\frac{v_\phi}{v_{\rm dis}}\right)^4 \gg 1 ,
\label{q2}
\end{equation}
or, in terms of the long azimuthal angle $\Omega\tau_{\rm dis}=Q^{1/2}$,
\begin{equation}
\Omega\tau_{\rm dis} = \frac{1}{2}\left(\frac{v_\phi}{v_{\rm dis}}\right)^2 ,
\label{q3}
\end{equation}
where $v_\phi = \Omega r$ is the orbital velocity. Equation~(\ref{q3}) corresponds to equation~(\ref{ot}) above; divided by $2\pi$, it gives the number of orbits in one dissipation time for fixed $\Omega$.

Equation~(\ref{q2}) reveals a fourth-power dependence of $Q$ on the ratio $v_\phi/v_{\rm dis} \gg 1$. The factor of 1/4 in it derives from the 1/2 seen in equation~(\ref{a11}) which also gives the same relation for $\ell = r v_\phi$ and $\Delta\varepsilon = v_{\rm dis}^2$ since from equation~(\ref{a3}),
\begin{equation}
\Delta \varepsilon = \left(\frac{{\cal L}}{\mu}\right)\tau_{\rm dis} = \epsilon\,\tau_{\rm dis}\, ,
\label{dvare}
\end{equation}
whereas from equations~(\ref{a1}) and~(\ref{q1}),
\begin{equation}
v_{\rm dis}^2 \equiv\epsilon\,\tau_{\rm dis}\, . 
\label{vare}
\end{equation}

\subsection{Integral Length}\label{IL}

The remaining scale in our problem, the integral length scale $l$ \citep{wan02}, derives from the above scales. For $Re=1$, we find that
\begin{equation}
l \equiv \frac{v_{\rm dis}^3}{\epsilon} = 
v_{\rm dis}\tau_{\rm dis} 
\, .
\label{il1}
\end{equation}
This $l$ is not the small length scale $\delta_{\rm dis}$ over which energy is dissipated;\footnote{Defined as $\delta_{\rm dis}\equiv v_{\rm dis}/\Omega$, the small length over which energy dissipation takes place is then found to be $\delta_{\rm dis} = r/(4Q)^{1/4}\ll r$ or, equivalently, $\delta_{\rm dis} = l/Q^{1/2}\ll l$.} it is the observable macroscopic length scale of the bulk kinetic energy of the bulges, some of which will be transferred to the much smaller dissipative scales $\sim\delta_{\rm dis}$ over times comparable to $\tau_{\rm dis}$. Its importance lies in the fact that the dissipation rate $\epsilon$ is primarily determined at this length scale via equation~(\ref{il1}), and not by the corresponding microscale $\delta_{\rm dis}$ of the ``turbulent'' regime \citep{geo13}. 

Using equations~(\ref{q2}), (\ref{q3}), and~(\ref{il1}), we find that
\begin{equation}
\left(\frac{l}{r}\right)^2 = \frac{1}{2}\Omega\tau_{\rm dis} 
\, ,
\label{il2}
\end{equation}
and that
\begin{equation}
\frac{l}{r} = \frac{1}{2} \left(\frac{v_\phi}{v_{\rm dis}}\right) = \left(\frac{Q}{4}\right)^{1/4} .
\label{il3}
\end{equation}
Perhaps a simpler interpretation derived from equation (\ref{il2}) (divide both sides by $\pi$) is that $l^2$ is the cumulative area $k(\pi r^2)$ that will be swept by the radius vector of a body after $k$ orbits taking place over time $t=\tau_{\rm dis}$.

\subsection{Damping Rate}

The damping rate $\gamma$ (dimension 1/time) of a wave-like perturbation on the surface of an incompressible fluid was derived by \cite{lan87} in their study of gravity waves of amplitude $A$, wavelength $\lambda \gg A$, and frequency $\omega \gg \nu/\lambda^2$. Their calculation appears to differ from above in two subtle respects: (a) \cite{lan87} define $\gamma$ as the coefficient of the decay of the amplitude $A$, not of the energy $\Delta\varepsilon$; and (b) they purport to calculate dissipation of the total mechanical energy, not only of the kinetic energy. 

Concerning difference (a), a relation between $\gamma$ and our $\tau_{\rm dis}$ is obtained by comparing the decay of the damped wave's energy $\Delta\varepsilon$ at any time $t$, viz.
\begin{equation}
\exp(-2\gamma t) = \exp(-t/\tau_{\rm dis})
\, ,
\nonumber
\end{equation} 
so that the damping rate of the amplitude $A$ is
\begin{equation}
\gamma = \frac{1}{2\tau_{\rm dis}}
\, .
\label{gamma}
\end{equation} 

Difference (b) above is more subtle because it does not seem to affect the scales of the problem; for example, using our notation and differentiating $\Delta\varepsilon\propto\exp(-2\gamma t)$ with respect to $t$, we find from the definition of $\gamma$ and equation~(\ref{gamma}) that
\begin{equation}
\frac{\epsilon}{2\Delta\varepsilon}\equiv\gamma = \frac{1}{2\tau_{\rm dis}} ~\Longrightarrow ~{\rm Equation~(\ref{dvare})} ,
\nonumber
\end{equation} 
so there is no difference between the two results. The reason is that despite the discussion preceding equation~(25.3) in \cite{lan87}, the energy they used is actually one-half of the mechanical energy of the perturbation, so the kinetic energy of the wave was actually used in their calculation as well.

\subsection{Remarks on Protostellar Disks}

We note that Kepler's third law was not used in the above calculations, so $\Omega$ was not assumed to necessarily be the equilibrium value, which is also fitting for the variational principle used in \S~\ref{free} above. In both cases, however, the bodies obey the two quasistatic conditions~(\ref{cond}) and~(\ref{condt}) or, equivalently, that $Q \gg 1$.

The above dissipation time $\tau_{\rm dis}$ should be accounted for in a planetary or satellite system after the gaseous accretion disk has dispersed because torques from the disk are expected to interfere in the early evolution of these bodies. Most protostars ($\sim$90\%) lose their inner disks after about 3-8 Myr \citep{hai01,hil08}, although some young stars apparently lose them within the first 1 Myr of their lifetimes and some older stars are found with inner disks after about 8-16 Myr. These timescales are shorter than the times over which terrestrial planet formation was completed in our solar system \citep[30-100 Myr;][]{wad00}. Owing to the soft dependence of $\tau_{\rm dis}/P$ on $Q^{1/2}$ seen in equation~(\ref{a8}), all of the above times are longer than the dissipation times $\tau_{\rm dis}$ of interacting solar subsystems, so there is ample time available for the solar nebula and gaseous protosatellite disks to disperse; and for the few (usually 4-7) developing massive cores to complete their accretion processes, differentiate themselves from their surroundings \citep{wad00}, and begin their next phase of quasistatic evolution driven by their mean tidal field and in the absence of other external torques. What occurs in this latter phase and the fate of the mean tidal field itself are the subjects of Appendix~\ref{alast}.

\section{Landau Damping of Tidal Waves near Mean-Motion Resonances}\label{alast}

\cite{gol65} studied local mean-motion resonances (MMRs) between pairs of satellites and found that the resonant configurations are not disturbed by tidal forces. This treatment confirmed that the tidal field created by the massive bodies in each subsystem seems to be absent when the bodies are near MMRs; but these local calculations did not provide a reason for the absence of the field. Nor could they, because MMRs are a global phenomenon that takes over the entire subsystem. \cite{gol65} could not imagine that the underlying field is nowadays severely weakened or dispersed altogether, so he hypothesized that the resonant bodies may regulate the transfer of angular momentum in ways that maintain the resonant configurations. Of course, this cannot be the case; the results in Appendix~\ref{n3} show that each body, resonant or not, receives and distributes angular momentum based on the conservation of the total amount and the small dissipation rate. So no body is capable of regulating transfer, although the most massive body will be perturbed much less solely because of its large inertia. Thus, we thought that this body is responsible for laying out the resonant structure of the subsystem globally, just as it provides a large part of the tidal field for its near-neighbors. It became apparent that when other massive bodies encountered principal MMRs of the most massive body, they would no longer contribute to the mean tidal field that they helped create in the first place, which, in turn, would be severely damped. Once the mean field (the collective mode of radial oscillations) was damped so, there was no mechanism to get it back. Minor bodies could not exchange angular momentum efficiently, thus they would also relax in nearby global MMRs sooner or later.\footnote{In our planetary system, the two largest by far deviations from nearby global MMRs of Jupiter occur in the terrestrial-planet subsystem: tiny Mars ($P=1.88$ yr) is distinctly short (by $-4.8$\%) of the 1:6 MMR and Venus ($P=0.615$ yr) is distinctly wide (by $+3.8$\%) of the 1:20 MMR. We entertain the thought that the tiny planet was pulled inward of this 1:2 local MMR with Earth by its two massive neighbors after the tidal field of the gaseous giants had dissipated away. Robbing Mars of its angular momentum may be what allowed Venus and Earth to move wide of their MMRs.}

Because of the above picture, we sought an explanation of the phenomenon in Landau damping \citep{lan46}, an analogous effect that takes place in electrons in a plasma. Gravitational Landau damping (LD) has already been applied to stellar systems \citep{lyn62,bin87,kan98,voo03}, but not to the few-body (4-7) systems that we envision. Thus, the historical trend in calendar time shows dramatic leaps from $10^{23}$ electrons in the 1940s, down to $10^{11}$ galaxy stars in the 1960s, and down again to 4-7 (extra)solar-system bodies nowadays. But there is no element in the derivation of LD that requires a large number of particles (furthermore, the fundamental assumption of a collisionless system is certainly satisfied by few bodies as opposed to $10^{11}$ stars or $10^{23}$ electrons). All that is required is a confinement mechanism, whether this be ionic Coulomb attraction in a plasma, or central gravitational attraction in a galaxy or in a few-body system. The first astrophysical studies made the connection between stellar systems and electronic plasmas because of the large numbers of ``particles'' involved \citep[also LD operates only at wavelengths that are stable to the Jeans instability;][]{jea02,tri04}; and they discovered that very small regions of the phase space of stellar systems contain the important particles (the so-called ``resonant'' particles) with speeds comparable to the phase velocity $v_{\rm ph}$ of the tidal wave. In retrospect, this must have been a surprise, as our reduction of the analysis to just 4-7 particles is also likely to be seen.

A complete satisfactory physical interpretation of LD was lacking until recently, although the outcome is no longer disputed. In plasmas, LD has been verified experimentally \citep{dov05,che19} and by simulations \citep{kli17}; it is also used to stabilize electron beams in accelerators \citep[][and references therein]{her14}. Recently, in the tradition of \cite{daw61}, the works of \cite{ryu99} and \cite{wes15} gave clear descriptions of LD using only real variables and their derivations make the physics behind the damping mechanism of the mean field much better understood. Further detailed descriptions using complex variables can be found in influential books on plasma physics \citep{lan81,sti92,bit04,bel06,fit15}; although such mathematical treatments may obscure to some extent the physics behind LD.

The damping mechanism in plasmas and stellar systems opearates as follows. Resonant particles gain energy from the mean field and become nonresonant, i.e., they move at speeds substantially higher than the phase velocity $v_{\rm ph}$ of the mean wave. Then other slower-moving particles become resonant and they gain energy from the field. The process continues until the field is robbed of its energy and dissipates away. This mechanism cannot work in exactly the same fashion in few-body systems because of the small number of ``particles'' involved. Instead, the mean field is weakened every time a massive body becomes precisely resonant (i.e., it ``levitates'' at the top of a wave crest) and the mean tidal field disappears altogether when the few major bodies all end up near resonances where they no longer support collective tidal interactions.

In what follows, we adopt the treatments of the linear LD by \cite{tri04} and \cite{fit15}, two resources providing clear physical insights, and we customize their analyses to the few-body gravitating systems of interest. We provide four theoretical derivations related to gravitational LD that are illuminating despite the mild use of complex variables; they complement nicely the real-value calculations recommended above \citep{daw61,ryu99,wes15}. First, we derive the characteristic screening length (analogous to the plasma Debye length) for few-body systems (\S~\ref{SL1}). Second, we verify that this screening length is formally precise for few-body systems, and we quantify the gravitational  Landau damping rate for the \cite{tri04} Jeans-stable waves (\S~\ref{SL2}). Third, we show a crucial elementary proof \citep{sti92,fit15} that bodies near the phase speed of such a wave will interact strongly with the wave, thus they are the ones participating in substantial energy exchanges and causing linear LD (\S~\ref{SL3}). Fourth, we describe the longitudinal oscillations of a single body initially in phase with the tidal wave and trapped in a potential trough of the decaying tidal field (\S~\ref{SL4}). Finally, we close with an application of the results to two important four-body subsystems in our solar system (\S~\ref{SL5}).

\subsection{Hill Radius and Jeans Wavenumber}\label{SL1}

Gravitational LD operates at short wavelenths $\lambda = 2\pi/k$, where $k$ is the wavenumber. The question is how short. With an eye on LD in stellar systems, \cite{bin87} determined the condition that $k > k_{\rm J}$ for standing waves to be necessarily damped (there are no travelling waves in the system),\footnote{They also perpetuated a common misconception that linear LD results from singularities in Landau's integrals. Here we pass on this issue and point, once again, to the calculations that did not use complex variables; there are no singularities in any of them.} where $k_{\rm J}$ is the critical Jeans wavenumber defined by the equation
\begin{equation}
k_{\rm J}\equiv\frac{\Omega_{\rm J}}{\sigma}
\, ,
\label{H1}
\end{equation}
where $\Omega_{\rm J}$ is the gravitational (Jeans) frequency and $\sigma$ is the velocity dispersion of stars.

For the few-body systems of interest, there is no predefined Jeans wavelength although we know empirically that the systems are dynamically stable, so they are not in any danger of suffering the dynamical Jeans instability. We need however to determine a cutoff value akin to $k_{\rm J}$. We proceed as follows. In plasma physics, the Debye radius is used to determine the volume inside which the field of one electron dominates relative to the mean field produced by all electrons. In our case, an analogous screening length is the Hill radius $h$,\footnote{{\tt https://en.wikipedia.org/wiki/Hill\_sphere}} that is
\begin{equation}
h = r \left(\frac{m}{3M}\right)^{1/3}
\, ,
\label{H2}
\end{equation}
where $r$ is the orbital radius, $M$ is the central mass, and $m$ is the mass of a body. Although not a constant, $h$ is a fair description of the sphere of gravitational influence around individual orbiting bodies.

It turns out that the above two scales are reciprocal. To show this, we need to redefine our concepts of Jeans frequency and velocity dispersion for few bodies with $m\ll M$ in Keplerian orbits about central mass $M$. We adopt the usual Keplerian orbital parameters, i.e., $\Omega^2=GM/r^3$ and $v_{\phi}^2 = GM/r$, where $G$ is the gravitational constant. The Keplerian orbital frequency is naturally the {\it de facto} Jeans frequency in this case, i.e., 
\begin{equation}
\Omega_{\rm J}^2 = \frac{GM}{r^3}
\, .
\label{H3}
\end{equation}
We also use the radial derivative of the $v_\phi^2$ given above, i.e., $2 v_{\phi}|\Delta v_r| = (GM/r^2)|\Delta r|$, $\sigma = |\Delta v_r|$, and $|\Delta r|=2h$. Then, combining these equations, we find that
\begin{equation}
\sigma = \Omega_{\rm J}h
\, ,
\label{H4}
\end{equation}
and then equation~(\ref{H1}) shows that
\begin{equation}
k_{\rm J} = \frac{1}{h}
\, ,
\label{H5}
\end{equation}

Another important quantity is determined when we transform the Jeans frequency that dictates the zeroth-order tidal field to a corresponding Hill frequency $\Omega_{\rm H}$ local to the individual bodies. Combining equations~(\ref{H2}) and (\ref{H3}), we find that
\begin{equation}
\Omega_{\rm H}^2\equiv \frac{Gm}{h^3} = 3\,\Omega_{\rm J}^2
\, .
\label{H6}
\end{equation}
As will be seen below, the factor of 3 is significant. For later reference, $h=0.355$ AU and $P_{H}\equiv 2\pi/\Omega_{\rm H} = 6.855$ yr for Jupiter in our planetary system now; and $h=31.72$ Mm and $P_{H} = 4.131$ d for Ganymede in Jupiter's satellite subsystem now.

\subsection{Jeans Instability and Landau Damping}\label{SL2}

The above relations are not order-of-magnitude estimates from dimensional analysis. The equations provide a precise description of the fundamental parameters that appear in the dispersion relation and the Landau damping rate for few-body systems. The exact same parameters have also been derived for a stellar system by \cite{tri04} (hereafter TEvS) in a fundamental piece of work that has been flying under the radar of the astronomical community for years. In particular:
\begin{itemize}
\item[(a)] TEvS considered a uniform ``infinite'' self-gravitating collection of masses with uniform density $\rho$, in which case the local Hill frequency $\Omega_{\rm H}$ is defined by the equation
\begin{equation}
\Omega_{\rm H}^2\equiv 4\pi G\rho 
\, .
\label{H7}
\end{equation}
\item[(b)] This idealized system contains two species of particles with masses $m$ and $M\gg m$. To rewrite $\Omega_{\rm H}^2$ as a ``global'' quantity, we imagine a spherical volume of radius $r$ containing a mass $M$ (smaller masses $\sim m$ are neglected) with mean density $\rho = 3M/(4\pi r^3)$, in which case we obtain
\begin{equation}
4\pi G\rho =  \frac{3 G M}{r^3} = 3\Omega_{\rm J}^2
\, ,
\label{H8}
\end{equation}
or $\Omega_{\rm H}^2 = 3\Omega_{\rm J}^2$, which is the same as equation~(\ref{H6}) for few-body systems. Now it becomes obvious why we used here the symbol $\Omega_{\rm H}$ for the local Hill frequency (TEvS call it $\Omega$), just as we did in \S~\ref{SL1} above. The need for radius $r$ to be taken around a mass $M$ stems from the peculiarities of this infinite uniform self-gravitating model (any mass $M$ can be a central mass in its vicinity).
\item[(c)] The linear stability analysis of this Jeans model also establishes a local ``Debye'' length scale 
\begin{equation}
D\equiv \frac{v_{\rm T}}{\Omega_{\rm H}},
\label{deb}
\end{equation}
which TEvS call the Debye-Jeans radius, although they point out incorrectly that this $D$ is not related to screening (a minor oversight that neglects the role of the Hill radius in gravitating bodies). Here $v_{\rm T}$ is the thermal velocity of fast particles belonging to the $m$-species in 3 dimensions. In one dimension, the velocity dispersion $\sigma$ will then be
\begin{equation}
\sigma^2 = v_{\rm T}^2/3\, ,
\label{svt}
\end{equation}
and then we find that
\begin{equation}
D \equiv \frac{v_{\rm T}}{\Omega_{\rm H}} = \frac{\sqrt{3}\sigma}{\sqrt{3}\Omega_{\rm J}} = \frac{1}{k_{\rm J}} = h .
\label{H9}
\end{equation}
Thus, the precise correspondence between parameters in the two models (\S~\ref{SL1} and \S~\ref{SL2}) is formally established.
\item[(d)] Collisions between heavy and light particles must be included in the TEvS model, otherwise the number of particles is not conserved. On the other hand, few-body systems are collisionless in the long term (some ejections of low-mass bodies by the massive bodies are expected in early evolution); thus, for our application, we reduce the equations of TEvS to the limit of zero collision frequency ($\nu\to 0$).
\item[(e)] The TEvS dispersion relation in the limit of $\nu\to 0$ reads
\begin{equation}
(k\, v_{\rm T})^2 - \Omega_{\rm H}^2\left[ 1 - J(\beta) \right] = 0
\, ,
\label{H10}
\end{equation}
where $k$ is the wavenumber and 
\begin{equation}
\beta\equiv(\omega + {\rm i}\nu)/(k\, v_{\rm T})\xrightarrow{\nu\to 0} \frac{\omega}{k\, v_{\rm T}}\, ,
\label{nu_cond}
\end{equation}
of a mode with frequency $\omega$. The function $J(\beta)$ is given by
\begin{equation}
J(\beta) \equiv \frac{\beta}{\sqrt{2\pi}}\int_{-\infty}^\infty\frac{\exp(-x^2/2)}{\beta - x} dx
\, ,
\label{H11}
\end{equation}
where, in our case, $x=v_{r}/v_{\rm T}$, with the asymptotic behavior
\begin{equation}
J(\beta) \approx -{\rm i}\sqrt{\frac{\pi}{2}}\beta, ~{\rm for}~|\beta|\ll 1
\, .
\label{H12}
\end{equation}
In equation~(\ref{H11}), the denominator $\beta -x$ is generally not singular owing to the presence of the collisional term $+{\rm i}\nu$ (equation~(\ref{nu_cond})). We distniguish two cases in the dispersion relation~(\ref{H10}):
\end{itemize}

{\sc Jeans Instability.}---For $|\beta|\gg 1$, when collisions are retained (that is, for $\beta\equiv(\omega + {\rm i}\nu)/(k\, v_{\rm T})$ in equation~(\ref{nu_cond})), equation~(\ref{H11}) can be integrated along the real axis. The dispersion relation takes the $\nu$-dependent form of equation (19) in TEvS. In the limit of $\nu\to 0$, it takes the asymptotic form
\begin{equation}
\omega^2 = 3(k\, v_{\rm T})^2 - \Omega_{\rm H}^2
\, .
\label{H13}
\end{equation}
Evidently, the two-species model exhibits the classical Jeans instability for long-wavelength modes ($k\ll k_{\rm J}$), provided that the characteristic ``sound speed'' $c_{\rm s}$ is defined as 
\begin{equation}
c_{\rm s}\equiv\sqrt{3}v_{\rm T}.
\label{cs}
\end{equation}
Relative to the few-body system of \S~\ref{SL1}, the two-species model then has
\begin{equation}
v_{\rm T}^2 = 3\sigma^2, ~
c_{\rm s}= 3\sigma, ~ D = h, ~{\rm and}~ \Omega_{\rm H}^2 \equiv 4\pi G\rho = 3\Omega_{\rm J}^2 .
\label{H14}
\end{equation}

{\sc Landau Damping.}---For $|\beta|\ll 1$ and for short wavelengths $k D > 1$, equations~(\ref{H10}) and~(\ref{H12}) combine to give
\begin{equation}
{\rm Re}(\omega)=0, ~~{\rm Im}(\omega)=\sqrt{\frac{2}{\pi}}\, k\, v_{\rm T}\left( 1 - k^2 D^2 \right) < 0
\, ,
\label{H15}
\end{equation}
in the limit of $\nu\to 0$. These are the Landau modes and they are all damping since ${\rm Im}(\omega) < 0$ for $k D > 1$. For wave amplitudes $\propto \exp(-\gamma t)=\exp({\rm Im(\omega)}t)$, the damping rate ($\gamma > 0$) is 
\begin{equation}
\gamma = \lvert {\rm Im}(\omega) \rvert = \sqrt{\frac{2}{\pi}}\, k\, v_{\rm T}\left( k^2 D^2 - 1 \right), ~~ (kD > 1)
\, .
\label{H16}
\end{equation}
For a few-body system ($D=h$), the damping rate $\gamma$ takes the form 
\begin{equation}
\gamma = \sqrt{\frac{2}{\pi}}\, \Omega_{\rm H} (k h)\left( k^2 h^2 - 1 \right), ~~ (k > k_{\rm J})
\, .
\label{H16}
\end{equation}
Thus, in this model, the damping rate is proportional to the local Hill frequency $(Gm/h^3)^{1/2}$ (equation~(\ref{H6})); whereas in the TEvS model, $\gamma$ is proportional to the local Jeans frequency $(4\pi G\rho)^{1/2}$ (equation~(\ref{H7})). In both models, waves with very short wavelengths ($k \gg k_{\rm J}$ or $kh\gg 1$) are damped at much higher rates ($\gamma \propto k^3$). On the other hand, waves with $kh \gtrsim 1$ and wavelengths
\begin{equation}
\lambda\lesssim 2\pi h\, ,
\label{Lwave}
\end{equation}
tend to persist for the longest times. In \S~\ref{SL5}, we describe an application of this result to the gaseous giant planets in our solar system and the Galilean satellites of Jupiter.

\subsection{Bodies Interacting with the Collective Field}\label{SL3}

We imagine that a one-dimensional radial tidal field ${\cal E}(r, t)$ generated by a few massive gravitating bodies is described by the equation
\begin{equation}
{\cal E}(r, t) = {\cal E}_0(r) \exp\left[ {\rm i} (kr - \omega t) \right]
\, ,
\label{cw1}
\end{equation}
where ${\cal E}_0$ is the amplitude (dimension of acceleration), $k$ is the radial wavenumber, and $\omega$ is the frequency of the wave. Any of the major bodies in this field feels an acceleration $dv_r/dt$ due to the collective influence of the other massive bodies of the same form, viz.
\begin{equation}
\frac{dv_r}{dt} = {\cal E}_0(r) \exp\left[ {\rm i} (kr - \omega t) \right]
\, .
\label{cw2}
\end{equation}
In the absence of the field, a body initially at $r=r_0$ with initial radial velocity $v_r=v_{r0}$ will move to $r=r_0+v_{r0}\, t$ and we can introduce the initial conditions to the perturbation by substituting the zeroth-order solution into the exponential term of equation~(\ref{cw2}) \citep{sti92,fit15}, viz.
\begin{equation}
\frac{dv_r}{dt} = {\cal E}_0(r) \exp\left[ {\rm i} (kr_0 + (k\, v_{r0} - \omega) t) \right]
\, .
\label{cw3}
\end{equation}
Integrating in time, we find for the velocity $v_r$ that
\begin{equation}
v_r - v_{r0} = {\cal E}_0(r)\left[\frac{\exp({\rm i}kr_0)}{{\rm i}k}\right]\left[\frac{\exp\left[ {\rm i} k (v_{r0} - \omega/k) t) \right] - 1}{v_{r0} -\omega/k}\right]
\, .
\label{cw4}
\end{equation}
For initial radial velocities $v_{r0}$ of bodies that are close to the wave's phase velocity 
\begin{equation}
v_{\rm ph}=\frac{\omega}{k}
\, ,
\label{phasev}
\end{equation}
we resolve the indeterminate form in the last bracket of equation~(\ref{cw4}) by de L'Hospital's rule, and we find that
\begin{equation}
v_r - v_{r0} = {\cal E}_0(r)\left[\exp({\rm i}kr_0)\right]\, t\, , ~~(v_{r0}\to v_{\rm ph})
\, .
\label{cw5}
\end{equation}
Thus, bodies with velocities close to $v_{\rm ph}$ (resonant bodies) will be subjected to linear velocity perturbations that grow in time. They will lose energy to the wave or gain energy from the wave, and they are responsible for the overall damping of the wave when it occurs eventually. This explains why in all related calculations, the damping rate $\gamma$ depends on the negative slope of the distribution function evaluated at $v=v_{\rm ph}$ \citep[e.g.,][]{wes15}. But it does not explain why wave damping predominates wave growth as the perturbed bodies may gain or lose energy in their interactions with the wave depending on their phases.

More detailed considerations are needed in order to understand the damping of the mean field. Following the clear descriptions given by \cite{fit15} and \cite{wes15}, we make the following important points for plasma fields and then for tidal fields:
\begin{itemize}
\item[(a)] It is certainly not the case that slightly faster-moving bodies will lose energy and slightly slower-moving bodies will gain energy from the wave, as is commonly quoted. This misconception invalidates the analogy with the famous example of a surfer riding an ocean wave. Whether a resonant body will gain or lose energy depends on the phase of the wave upon energy exchange. In other words, a radially oscillating body trapped within its Hill radius with radial veclocity near the wave's phase velocity will rob the wave of some of its energy only if its oscillation is in phase with the wave (see \S~\ref{SL4} below). 
\item[(b)] The ``density'' perturbation generated by a displaced body is not in phase with the wave \citep{wes15}, so the initial wave cannot generate an initial distribution in which energy gain or loss by bodies is favored \citep{fit15}.
\item[(c)] Considering only resonant bodies starting with velocities $v \gtrsim v_{\rm ph}$, if they gain energy, they will move away from resonance; whereas if they lose energy, they will move closer to the resonant velocity $v_{\rm ph}$. The end result is that the latter bodies interact more efficiently with the wave and, on average, the field gains energy from bodies with $v \gtrsim v_{\rm ph}$. The opposite holds for bodies with $v \lesssim v_{\rm ph}$ for which the gainers are more efficient and the field is damped.
\item[(d)] In a Maxwellian radial velocity distribution (even an unusual one with just 4-7 bodies) or in any other distribution with a roughly similar (bell-shaped) profile, there will be more bodies with $v \lesssim v_{\rm ph}$; thus on average, the wave will have to push on most of them and it will be damped. It is for this reason that the {\it negative gradient} of the distribution function at $v=v_{\rm ph}$ determines the damping rate \citep{wes15}. 
\end{itemize}

We note however that items (c) and (d) above do not play an important role in few-body systems because few bodies have another mechanism available to them in order to cease contributing to the mean tidal field thereby undermining it to a great extent. We describe LD carried out by few gravitating bodies in \S~\ref{SL4} and \S~\ref{SL5} below.

Landau damping in gravitating systems \citep{bin87,tri04} has different origin than in electronic plasmas \citep{lan81,sti92,bit04,bel06}. Furthermore, in plasmas, the effect appears for eigenvalues with $|{\rm Re}(\omega)|/k \gg v_{\rm T}$ \citep{wes15}; whereas in stellar systems it appears for imaginary eigenvalues of the form $|{\rm Im}(\omega)|/k \ll v_{\rm T}$ and in the opposite limit, only for eigenvalues with $|{\rm Im}(\omega)| \gg |{\rm Re}(\omega)|$ and $kD\gg 1$ \citep{tri04}, or with $|{\rm Im}(\omega)| \gtrsim |{\rm Re}(\omega)|$ \citep{bin87}. These distinctions argue against using  the term``Landau damping'' for both types of systems. Using the term ``gravitational Landau damping'' (GLD) for astrophysical systems apparently resolves this issue.

\begin{figure}
\begin{center}
    \leavevmode
      \includegraphics[trim=0.1cm 0.1cm 0.1cm 0.1cm, clip, angle=0,width=8.5 cm]{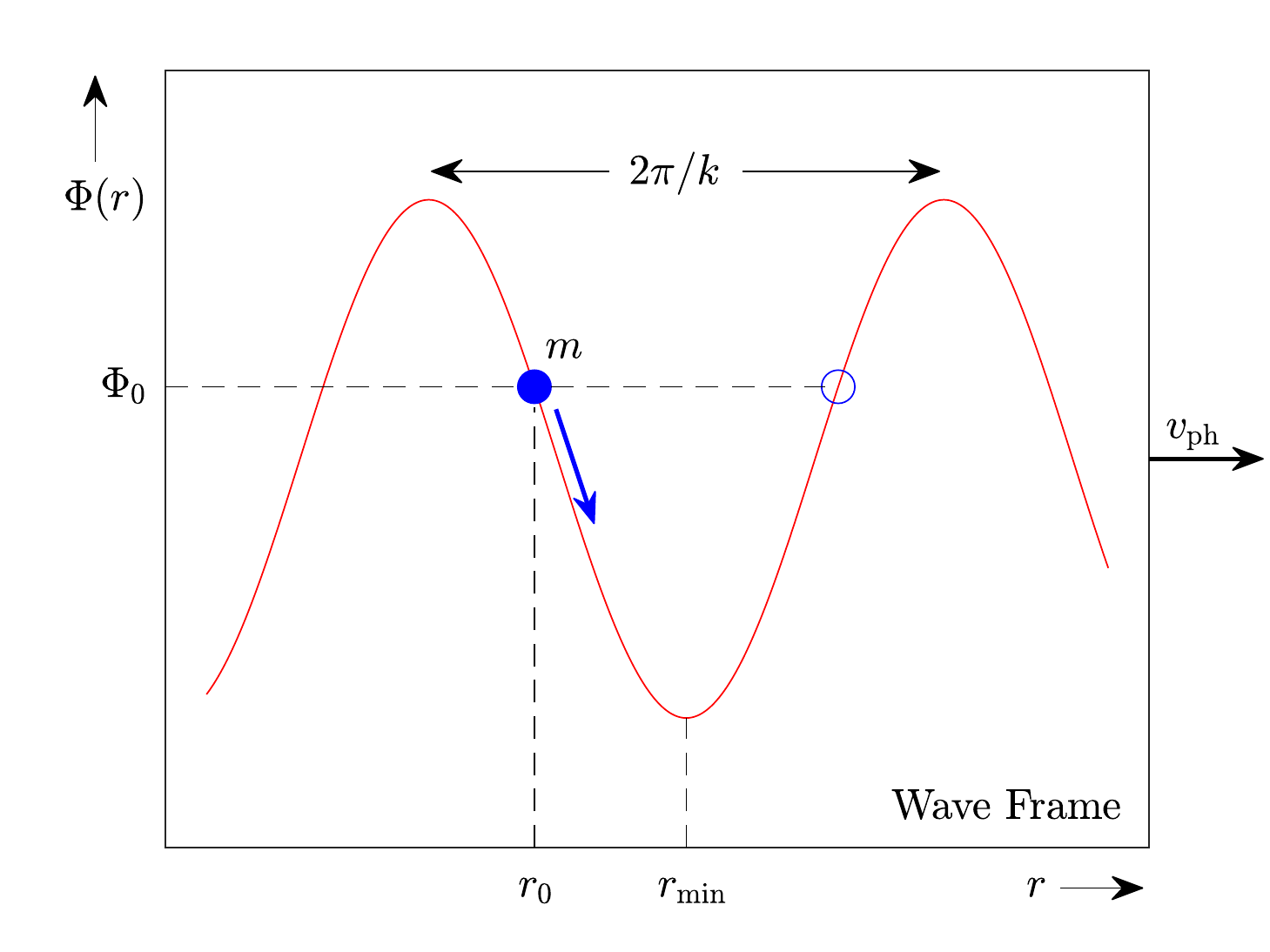}
\caption{Schematic diagram of a body of mass $m$ trapped in a trough of the potential $\Phi(r)$ due to the standing tidal wave ${\cal E}(r) = k\Phi(r)$ described in \S~\ref{SL3} above. The body starts at $r=r_0$ with relative velocity $v_r=0$ (in phase). As it is settling toward $r=r_{\rm min}$, it bounces back and forth radially between turning points of opposite phases, such as the points indicated by circles.
\label{wave1}}
  \end{center}
\end{figure}

\subsection{A Body Trapped in the Tidal Field}\label{SL4}

\subsubsection{Linear Regime}

Consider a resonant body, initially at rest at $r=r_0$, trapped in a trough of the mean potential $\Phi(r)=-\int{\cal E}(r)dr$ of a standing tidal wave, in a reference frame that moves with the phase velocity $v_{\rm ph}$ of the wave \citep[e.g.,][]{sti92}, as shown in Figure~\ref{wave1}. The turning points of the potential are specified by the value $\Phi_0$. The body will bounce around the potential minimum at $r=r_{\rm min}$ according to the harmonic oscillator equation
\begin{equation}
\frac{d^2r}{dt^2} = - k^2\Phi_0 \left(r - r_{\rm min}\right) ,
\label{har1}
\end{equation}
with $r(0)=r_0$, $v_{r0}=\frac{dr}{dt}(0)=0$, and solution 
\begin{equation}
r = r_{\rm min} - A\cos\left(\sqrt{\Phi_0}\,kt\right) ,
\label{har2}
\end{equation}
where the amplitude $A=\left(r_{\rm min}-r_0\right)$. Any small amount of dissipation $\gamma(dr/dt)$ (such that $\gamma/2\ll k\sqrt{\Phi_0}$) in equation~(\ref{har1}) will modify the amplitude $A$ in equation~(\ref{har2}) to $A_{\gamma} = A\exp(-\gamma t/2)$ and will drive $A_{\gamma}(t)$ toward zero; and if the body relaxes to $r=r_{\rm min}$, it will levitate there---i.e., it will keep moving with the wave without gaining or losing energy or angular momentum (for orbiting bodies). When most bodies in a few-body system relax near potential minima, then the mean field will be damped out. This is how the tidal field is weakened and finally is dispersed, when the relaxed major bodies no longer contribute to it. Obviously, this mechanism is not at all efficient in plasma or stellar systems, but it is ideal for the few-body systems considered here.

For the wave described by equation~(\ref{cw1}), the frequency of the bounce is $\omega_{\rm B}^2 = k^2\Phi_0 = k {\cal E}_0$. The characteristic period of the bounce then is $P_{\rm B}\propto 1/\sqrt{{\cal E}_0}$, i.e., it becomes very long as the wave amplitude ${\cal E}_0$ decays in time. In such a case, the body should always be found near $r=r_{\rm min}$ at times $t > \tau_{\rm dis}$ (long after the tidal wave has effectively dissipated). In an ironic twist, this observation rekindles a discussion of the Titius-Bode rule, hopefully for the last time (see \S~\ref{SL5} below). 

For the longest and slower-damped modes with $k\gtrsim k_{\rm J}$ (i.e., $k\approx 1/h$), then ${\cal E}_0 = \Omega_{\rm H}^2 h$ and the bounce frequency is $\omega_{\rm B} = \Omega_{\rm H}$. In this limit, the maximum period of the bounce is $(P_{\rm B})_{\rm max} = P_{\rm H}$, for which characteristic values were given in \S~\ref{SL1} for the radial movements of Jupiter and Ganymede, albeit using presently observed values; it turns out that $(P_{\rm B})_{\rm max}$ is $1/\sqrt{3}\approx 0.577$ of the orbital period of each body; and the corresponding wavelengths from equation~(\ref{Lwave}) are 2.233 AU and 199.3 Mm, respectively. These values will be used in the application of \S~\ref{wav} below.

In general, the settled bodies in a system in which the tidal field has been damped are not expected to be found all in phase because of the bouncing around in their potential troughs that preceded their settling. We have experienced this situation first-hand in the work of \cite{gol65} who found only 7 pairs of satellites of the gaseous giants having related phase angles. In expanding the search for mean-motion resonances in (exo)planetary subsystems, we need to search, not only for approximate scalings of the orbital-period ratios and phase angles, but for spatial wavelength-dependent scalings as well. To do the latter part, first we need to obtain an estimate of the longest wavelength of the mean tidal field long gone; but this may not be such a difficult task, as is demonstrated in \S~\ref{SL5} below.

\subsubsection{Nonlinear Regime}

As the tidal field is being damped by the participating bodies, its amplitude ${\cal E}_0(t)$ decreases in time. The frequency of the bounce $\omega_{\rm B}(t)$ also decreases and the oscillations of a body such as $m$ in Figure~\ref{wave1} take longer times $P_{\rm B}(t)=2\pi/\omega_{\rm B}$. Let ${\cal E}_0(0)$, $\omega_{\rm B}(0)$, and $P_{\rm B}(0)$ be the initial values at $t=0$. At early times, the damping of the wave proceeds according to Landau's linear theory, that is ${\cal E}_0(t)\propto\exp(-\gamma t)$, where $\gamma\ll\omega_{\rm B}$ is the (small) damping rate. At later times, when $\omega_{\rm B}(t)\sim\gamma$, nonlinear oscillations appear in the the plasma variables \citep{one65,arm67}. On the other hand, we would like to know how fast damping develops on average at such later times in few-body systems.

If present in few-body systems, the nonlinear oscillations will have frequencies similar to $\omega_{\rm B}(t)$ and they will be superposed to the overall decaying amplitude \citep{one65,fit15}. For 
\begin{equation}
\omega_{\rm B}(t)=2\pi\gamma ~~({\rm i.e., for}~ P_{\rm B}(t)=1/\gamma), 
\label{neweq}
\end{equation}
the equations for the time-dependent amplitude and the bounce frequency take the forms
\begin{equation}
y = \exp(-\tau\sqrt{y}) ,
\label{non1}
\end{equation}
and
\begin{equation}
x = \exp\left(\frac{\tau\sqrt{y}}{2}\right) = \exp\left(\frac{\tau}{2x}\right),
\label{non2}
\end{equation}
respectively, where $\tau = t/P_{\rm B}(0)$, $y = {\cal E}_0(t)/{\cal E}_0(0)$, and $x = P_{\rm B}(t)/P_{\rm B}(0)$. Eliminating time between these two equations, we get back the relation $y=1/x^2$ (or $\omega_{\rm B}^2/{\cal E}_0=k = {\rm const.}$) valid at all times. The solutions for $y$ and $x$ are given in terms of the Lambert $W$ function \citep{cor96,val00} with argument $\xi = \tau/2$, viz.
\begin{equation}
y = \left(\frac{W(\xi)}{\xi}\right)^2 = e^{-2W(\xi)} ,
\label{non3}
\end{equation}
and
\begin{equation}
x = \left(\frac{W(\xi)}{\xi}\right)^{-1} = e^{W(\xi)} .
\label{non4}
\end{equation}
The intermediate time $\tau_\star$ (or $\xi_\star$) at which the equations are exactly valid depends on the magnitude of $\gamma$ since $P_{\rm B}(t)$ was set equal to $1/\gamma$. Let $\Upsilon$ be the dimensionless value corresponding to the value of $\gamma$, that is, let $\Upsilon = \gamma P_{\rm B}(0)$. Then 
\begin{equation}
\Upsilon_\star = \frac{P_{\rm B}(0)}{P_{\rm B}(t_\star)} = \frac{1}{x_\star} = e^{-W(\xi_\star)} ~\Longrightarrow ~W(\xi_\star) = -\ln\Upsilon_\star\, ,
\label{def1}
\end{equation}
where $\xi_\star = \tau_\star/2$ and $\tau_\star = t_\star/P_{\rm B}(0)$. This Lambert function can be readily inverted and the principal branch gives
\begin{equation}
\xi_\star = -\frac{\ln\Upsilon_\star}{\Upsilon_\star} ~\Longrightarrow ~ \tau_\star =  -2 \frac{\ln\Upsilon_\star}{\Upsilon_\star}.
\label{def2}
\end{equation}
For example, for $\Upsilon_\star = 0.25$, the intermediate time is $\tau_\star = 11.09$, corresponding to 11 bounces with period $P_{\rm B}(0)$.

\subsubsection{Asymptotic Courses}

(a) At early times ($\xi < 1$) and for $\gamma < 1/P_{\rm B}(0)$ (i.e., $\Upsilon < 1$), the series expansions of $y(\tau)$ and $x(\tau)$ are
\begin{equation}
y(\tau) = 1 - \tau + \tau^2 - 25\tau^3/24 + {\cal O}\left(\tau^4\right),
\label{non5}
\end{equation}
and
\begin{equation}
x(\tau) = 1 + \tau/2 - \tau^2/8 + \tau^3/12 + {\cal O}\left(\tau^4\right).
\label{non6}
\end{equation}
Amplitude decay starts out with a steep slope of $dy/d\tau = -1$ and at later times, the slope approaches zero from below quite fast ($dy/d\tau(10)=-0.00803$, $y(10)=0.0704$). Bounce period stretch starts out with a slope of $dx/d\tau = 1/2$ and at later times, the slope approaches zero from above slowly ($dx/d\tau(10)=0.215$, $x(10)=3.77$). \\

\noindent
(b) At late times ($\xi \gg 1$) and for much smaller values of $\gamma\sim 1/P_{\rm B}(t)$ (i.e., $\Upsilon\sim 1/x$), the Lambert $W$ function can be approximated by 
\begin{equation}
W(\xi)\sim \ln\left(\frac{\xi}{\ln\xi}\right) ,
\label{rough}
\end{equation}
to leading order on its principal branch \citep{cor96}, and equations~(\ref{non3}) and~(\ref{non4}) can be approximated by the coarse asymptotic forms
\begin{equation}
y\sim \left(\frac{\ln\xi}{\xi}\right)^2 ,
\label{non7}
\end{equation}
and
\begin{equation}
x\sim \left(\frac{\ln\xi}{\xi}\right)^{-1} ,
\label{non8}
\end{equation}
respectively. For comparison purposes, $dy/d\tau\approx -0.00785$, $y\approx 0.104$, $dx/d\tau\approx 0.118$, and $x\approx 3.11$ for $\xi=5$ ($\tau=10)$. Since $\xi = \tau/2$, these equations describe the time dependence of the the amplitude and the bounce period at intermediate times; i.e., at the onset of the nonlinear regime, although it is known that the linear approximation continues to be valid well into this regime \citep{one65,arm67}.

\subsection{Signatures of Tidal Fields Long Gone}\label{SL5}

\subsubsection{Imprints}

According to the results of our study, major planets in our planetary system and massive moons in satellite subsystems moved around in their collective tidal fields until they got caught in potential troughs where they settled near potential minima and contributed to the damping of the field. Damping occurred because most, if not all, bodies developed radial speeds equal to the phase velocity of the longitudinal wave. In such a levitating configuration, tidal interactions ceased and the wave was severely suppressed. In such a case, there must be imprints left over in the currently settled orbits of major bodies, signatures of a tidal dissipative evolution that took place in the distant past. Some imprints were found by \cite{gol65} in the phase angles of some resonant satellite pairs and in the Laplace phase of the three inner Galilean moons. Below we pursue additional imprints in the wavelengths of long-gone tidal fields.

\subsubsection{Wavelengths}\label{wav}

We search for the most obvious imprints of such evolutions in solar-system subsystems, those related to the wavelength of the tidal field. Once again, the mere premise of this search is at odds with the phenomenology surrounding the empirical Titius-Bode (TB) rule (while, at the same time, the results confirm the conclusions of \cite{las00} and \cite{chr17}---planets settled at locations in which nearest neighbors were no longer interacting with one another). As we pointed out in the past, the orbital radii of the 3 innermost planets and the 3 outer gaseous giants are {\it obviously} in arithmetic progression, in clear contradiction with the geometric progression of the TB rule \citep{chr17}. This very old observation fits quite well in the present context of equidistant potential minima in the expired tidal field.

In order to search for radial regularities in the current orbits of solar-system bodies, we need to have some prior knowledge about the longest wavelength $\lambda$ of the long-gone tidal field. Equation~(\ref{Lwave}) is a suitable starting point, but this is not the regularity condition we seek for the following reason: Neighboring bodies cannot generally settle into adjacent potential minima (\S~\ref{SL4}) because they cannot both control the same Hill sphere. Therefore, nearest-neighboring bodies must be generally separated by at least two wavelengths of the tidal field. Thus, we define ${\cal S}_{\rm min}$, the minimum separation between adjacent bodies, by the equation
\begin{equation}
{\cal S}_{\rm min}\equiv 2\lambda = 4\pi h\, ,
\label{lam1}
\end{equation}
where $h$ is the Hill radius of the most massive body in the subsystem. Then, for the wavelengths of the tidal fields of the gaseous giant planets and the Galilean satellites around Jupiter given in \S~\ref{SL4}, we find that the minimum separations are
\begin{equation}
{\cal S}_{\rm min} \simeq 4.5~{\rm AU}, ~~({\rm Gaseous~Giants}) ,
\label{lam2a}
\end{equation}
and
\begin{equation}
{\cal S}_{\rm min} \simeq 0.4~{\rm Gm}, ~~({\rm Galilean~Moons}) ,
\label{lam2b}
\end{equation}
respectively. We apply, in turn, these ${\cal S}_{\rm min}$ estimates to the corresponding solar subsystems below.

\subsubsection{Gaseous Giants}

We consider the gaseous giant planets in our solar system. It is well-known that their orbital radii are $\approx$ 5, 10, 20, and 30 AU, respectively. The arithmetic progression that should have invalidated the empirical TB rule long ago is obvious in the last three radii. If these four massive planets are largely responsible for the damping of the collective wave during dissipative evolution in the past, then they must have finally settled near the bottoms of what used to be wave troughs of the standing tidal wave that pushed them around for a time. 

This is clearly confirmed by the present-day orbital radii of the gas giants. Using the 4.5 AU minimum separation (equation~(\ref{lam2a})), we find that relative to Jupiter ($r_{\rm Ju}=5.20$ AU), the outer three gaseous giants settled at about 2, 6, and 11 wavelengths away; the predicted radii are 
\begin{equation}
9.70, ~18.7, ~{\rm and}~ 30.0~ {\rm AU}; 
\label{plseq}
\end{equation}
to be compared with the actual semimajor axes of 9.58, 19.2, and 30.1 AU, respectively (relative deviations $< 3$\%). So Jupiter and Saturn are confirmed to be adjacent neighbors and Kepler's third law gives an orbital period ratio of $(9.70/5.20)^{3/2} \simeq 5/2$ with a relative deviation of only 2\%. (On the other hand, the precise ratio of orbital radii is 1.842 and Kepler's third law then gives a period ratio of $1.842^{3/2} = 5/2$ precisely.) It becomes obvious then that this is a pristine resonant subsystem with the four gaseous giants having settled (at increasing orbital periods) near the 1:1, 5:2, 7:1, and 14:1 MMRs of Jupiter.

\subsubsection{Galilean Moons}

Next we consider the four massive Galilean moons of Jupiter (Io, Europa, Ganymede, and Callisto) in some detail. Their orbital radii are
\begin{equation}
0.42, ~0.67, ~1.07, ~{\rm and}~ 1.88 ~{\rm Gm} , 
\label{seq0}
\end{equation}
respectively. This sequence has not been subjected to dubious numerological analyses in the past, so our estimates (and the physics behind them) are new and incomparable. Using the 0.4 Gm minimum separation for Ganymede (equation~(\ref{lam2b})), we find that Europa is 2 wavelengths inward and Callisto is 4 wavelengths outward of Ganymede. The precision of this orbital configuration is astounding by astronomical measures. It has not been quoted or discussed in the past because a physical model such as GLD of the tidal field was lacking.

On the other hand, Io appears to have settled at 3.25 wavelengths inward of Ganymede and its location reveals that it is adjacent to Europa. (The number of wavelengths is not an integer probably because Io was locked into the Laplace resonance early on.) Although not expected, this $\sim1\lambda$ separation is easily understood because, owing to their smaller masses, the Hill radii of Io and Europa are much smaller than that of Ganymede (by factors of 0.33 and 0.43, respectively). Thus, although these smaller moons are adjacent neighbors, their Hill spheres do not at all overlap.

Scaled to the orbital radius of Io ($r_{\rm Io}=421.7$ Mm), the orbital radii of the 4 Galilean moons are 
\begin{equation}
1, ~1.6, ~2.5, ~{\rm and}~ 4.5,
\label{seq}
\end{equation}
respectively. (Europa, the smallest moon, is a bit off of 1.5 in this scale for the reason noted in \S~\ref{dis}---it was locked into the Laplace resonance early on.) Thus, counting out by +0.5 from Io occupying past wave trough 1, the next three Galilean moons have settled very close to the potential minima of tidal-wave troughs 2, 4, and 8. This is how the Laplace resonance is realized in the spatial dimension of the long-gone tidal field, but only in conjunction with Kepler's third law which must be valid for the observed spatial layout to be confirmed as resonant: in particular, relative to the orbit of Io, the Keplerian period ratios are $1.6^{3/2} = 2.0$ for Europa and $2.5^{3/2} = 4.0$ for Ganymede.

Callisto, the outermost very massive moon,\footnote{Callisto is the third most massive moon in the solar system behind Ganymdede and Saturn's Titan. Its mass is 72.6\% of Ganymede's and 80.0\% of Titan's.} is famous for not participating in the 1:2:4 Laplace resonance of the innermost three moons and having to settle down to the 7:3 global mean-motion resonance relative to the most massive moon Ganymede \citep{mur99}. From the spatial sequence~(\ref{seq}), we get for Callisto and Ganymede $4.5/2.5 = 1.8$ and a period ratio of $1.8^{3/2}\simeq 7/3$ with a relative deviation of 3.5\%. (On the other hand, the precise ratio of orbital radii is 1.759 and Kepler's third law then gives a period ratio of $1.759^{3/2} = 7/3$ exactly.)
 
Finally, we note that Callisto could not have settled closer to Ganymede than $4\lambda$ as presently observed. Had it been settled at the $1\lambda$ or $2\lambda$ potential minima (radii 1.27 Gm and 1.47 Gm, respectively), the Hill spheres of the two major moons would overlap (at $2\lambda$, Callisto's Hill radius would be 0.25 Gm, causing overlap with Ganymede's Hill sphere). Although that would not have been the case were Callisto orbiting at $3\lambda$ (radius 1.67 Gm), where its Hill radius would be 0.28 Gm and, in addition, Callisto would be on the 2:1 global MMR of Ganymede, extending thus the Laplace chain to 4 moons. 

We believe that the prospect of being in this 2:1 global MMR is precisely what made the $3\lambda$ orbit unreachable to Callisto. There is ample evidence in the satellite subsystems of our solar system and in exoplanetary systems (Christodoulou \& Kazanas, in prep.) that the 1:2 global resonance is ``forbidden,'' unless it is a building block of a Laplace triple chain \citep[see, e.g., GJ 876;][]{riv10,mil18}. Also, quadruple Laplace chains (1:2:4:8 MMRs) do not appear to be stable, with the last arriving member (number 1 or 8) being pushed away from either side of the already-formed triple chain. Investigation of this interesting subject is not closed at this point \citep[see also][]{ger12, mar13}.


\label{lastpage}

\end{document}